\documentclass[aps,prx,reprint,letterpaper,superscriptaddress,longbibliography]{revtex4-2}
\usepackage{graphicx}
\usepackage{amsmath}
\usepackage{mathptmx}
\usepackage{bm}
\usepackage[colorlinks,linkcolor=blue,anchorcolor=blue,citecolor=blue,urlcolor=blue]{hyperref}
\makeatletter
\newcommand{\setcontactauthororder}[2]{%
  \begingroup
    \def\@mpfn{mpfootnote}%
    \c@mpfootnote=0\relax
    \setbox0=\hbox{%
      \frontmatter@footnote{Contact author: \href{mailto:#1}{#1}}%
    }%
    \setbox0=\hbox{%
      \frontmatter@footnote{Contact author: \href{mailto:#2}{#2}}%
    }%
  \endgroup
}
\makeatother
\begin{document}
\setcontactauthororder
  {weiku@sjtu.edu.cn}
  {vadim.grinenko@sjtu.edu.cn}
\title{Effective Ionic Valence and Local Magnetic Moment in Kagome Superconductors}
\author{Ruoshi Jiang}
\affiliation{School of Physics and Astronomy and Tsung-Dao Lee Institute, Shanghai Jiao Tong University, Shanghai 201210, China}
\affiliation{Department of Materials Science and Metallurgy, University of Cambridge, Cambridge CB3 0FS, United Kingdom}
\author{Zi-Jian Lang}
\affiliation{School of Physics and Astronomy and Tsung-Dao Lee Institute, Shanghai Jiao Tong University, Shanghai 201210, China}
\author{Yuzki Oey}
\affiliation{Materials Department, University of California Santa Barbara, Santa Barbara, California 93106, USA}
\author{Andrea Capa Salinas}
\affiliation{Materials Department, University of California Santa Barbara, Santa Barbara, California 93106, USA}
\author{Stephen D. Wilson}
\affiliation{Materials Department, University of California Santa Barbara, Santa Barbara, California 93106, USA}
\author{Yongwei Li}
\affiliation{Tsung-Dao Lee Institute, School of Physics and Astronomy, and State Key Laboratory of Micro-nano Engineering Science, Shanghai Jiao Tong University, Shanghai 201210, China}
\author{Ilya Shipulin}
\affiliation{Leibniz Institute for Solid State and Materials Research, D-01069 Dresden, Germany}
\affiliation{Tsung-Dao Lee Institute, School of Physics and Astronomy, and State Key Laboratory of Micro-nano Engineering Science, Shanghai Jiao Tong University, Shanghai 201210, China}
\author{Yiwen Zhang}
\affiliation{Tsung-Dao Lee Institute, School of Physics and Astronomy, and State Key Laboratory of Micro-nano Engineering Science, Shanghai Jiao Tong University, Shanghai 201210, China}
\author{Deng Hu}
\affiliation{Centre for Quantum Physics, Key Laboratory of Advanced Optoelectronic Quantum Architecture and Measurement (MOE),
School of Physics, Beijing Institute of Technology, Beijing 100081, China}
\author{Zhiwei Wang}
\affiliation{Centre for Quantum Physics, Key Laboratory of Advanced Optoelectronic Quantum Architecture and Measurement (MOE),
School of Physics, Beijing Institute of Technology, Beijing 100081, China}
\affiliation{Beijing Institute of Technology, Zhuhai 519000, China}
\author{Hans-Henning Klauss}
\affiliation{Institute for Solid State and Materials Physics, Technische Universit\"at Dresden, D-01069 Dresden, Germany}
\author{Zurab Guguchia}
\affiliation{PSI Center for Neutron and Muon Sciences CNM, 5232 Villigen PSI, Switzerland}
\author{Vadim Grinenko$^\dagger$}
\affiliation{Tsung-Dao Lee Institute, School of Physics and Astronomy, and State Key Laboratory of Micro-nano Engineering Science, Shanghai Jiao Tong University, Shanghai 201210, China}
\author{Wei Ku$^*$}
\affiliation{School of Physics and Astronomy and Tsung-Dao Lee Institute, Shanghai Jiao Tong University, Shanghai 201210, China}
\affiliation{Key Laboratory of Artificial Structures and Quantum Control (Ministry of Education), Shanghai 200240, China}
\affiliation{Shanghai Branch, Hefei National Laboratory, Shanghai 201315, People's Republic of China}

\begin{abstract}
In order to understand the unexpected similarity and the correlated behavior in kagome superconductor families AV$_3$Sb$_5$ (A = K, Rb, Cs) and ATi$_3$Bi$_5$ (A = Rb, Cs), we investigate the Hartree-scale local electronic structure of these systems.
Our result indicates that V and Ti ions are both of 2+ valence such that the corresponding itinerant carrier densities are similar, and the difference in electron count is instead reflected in their quantum fluctuating ionic magnetic moments.
However, due to the frustrated lattice geometry of these materials, such local moments are difficult to experimentally observe via standard probes.
For verification, we systematically introduce nonmagnetic Sn impurities to locally relieve the geometric frustration and experimentally demonstrate the existence of well-defined local magnetic moments via magnetic susceptibility and muon spin rotation or relaxation ($\mu$SR) measurements.
All experiments discover a systematic increase of magnetic susceptibility upon increasing nonmagnetic impurity level.
Our discovered ionic moments suggest a paradigm shift from the existing itinerant carrier-only picture to one incorporating strong correlation from local ionic spins.
The associated interatomic and local-itinerant correlations offer a solid ground for the emergence of the observed rich correlated behavior in this new family of superconducting materials.
\end{abstract}
\maketitle

\section{Introduction}

Materials with kagome lattice offer a fertile platform to study exotic quantum states of matter resulting from their frustrated geometry, novel correlation, and topological electronic structure.
Recently, the layered kagome metal $\rm AV_3Sb_5$ $\rm (A=K, Rb, Cs)$~\cite{Ortiz2019New} has attracted tremendous attention due to the rich physical behavior it hosts, including charge density waves (CDWs)~\cite{Kang2022Twofold, Chen2021Roton, Khasanov2022Time-reversal, Jiang2021Unconventional, Xiang2021Twofold, Ratcliff2021Coherent, Uykur2022Optical, Ortiz2021Fermi, Li2022Coexistence, Lou2022Charge-Density-Wave-Induced,Li2021Observation, Luo2022Electronic, Li2022Rotation, Nie2022Charge-density-wave-driven, Wu2022Charge, Yu2021Evidence, Park2021Electronic, Lin2021Complex, Denner2021Analysis, Tan2021Charge, Christensen2021Theory, Setty2021Electron, Feng2021Low-energy, Zhou2022Chern, Tazai2022Mechanism, Subedi2022Hexagonal-to-base-centered-orthorhombic, Liu2021Charge-Density-Wave-Induced, Shumiya2021Intrinsic, Wang2021Charge, Feng2021Chiral, Wu2021Nature, Mielke2022Time-reversal, Liang2021Three-Dimensional, Zheng2022Emergent, Wang2023Structure, Wang2021Distinctive, Yu2021Concurrence, Zhou2021Origin, Ge2023Anharmonic, Zhao2021Cascade,Guguchia2023Tunable, Gupta2022Two, Li2022No},
unconventional superconductivity~\cite{Ortiz2019New,Chen2021Roton,Yin2021Superconductivity,Du2021Pressure,Yu2021Unusual,Zhu2022Double-dome,Zhao2021Cascade,Song2021Enhancement,Duan2021Nodeless,Wang2021Charge,Mu2021S-Wave,Nakayama2022Carrier,Nguyen2022Electronic,Song2022Orbital,Salinas2023Electron-hole}, and anomalous Hall and Nernst effects~\cite{Yang2020Giant, Chen2022Anomalous, Mi2022Multiband, Yu2021Concurrence}.
Among these rich physics, the anomalous transport and magnetic behaviors demonstrate clear signatures of electronic correlation~\cite{Nguyen2022Electronic} beyond the standard paradigm of Fermi liquid.

Most recently, an isostructural new family of Ti-based kagome metals $\rm ATi_3Bi_5$ $\rm (A=Rb, Cs)$ has been found to display similar rich properties, such as double-dome superconductivity under pressure~\cite{Du2021Pressure,Nie2023Pressure}, electronic nematicity~\cite{Yang2022Superconductivity, Yong2022Nontrivial, Li2023Electronic, Guo2024Correlated}, and the nontrivial band topology~\cite{Yang2022Titanium,Yang2022Titanium-based,Yong2022Nontrivial}.
Given that the nominal electron count of this new family is three electrons per chemical unit less than that of the V-based family, rather distinct physical properties are expected due to the different Fermi surfaces~\cite{Yong2022Nontrivial, Ortiz2020CsV3Sb5}.
Surprisingly, with such a dramatically different electron count, these two families still display many similar \textit{unorthodox} physical properties.
A natural possibility is, therefore, a common mechanism of higher energy scale and emerged from it strong short-range correlations that surpass the standard paradigm of Fermi liquid.

As is well established in the studies of transition metal oxides~\cite{Anderson1972,Dagotto2005Complexity}, such higher-energy mechanism typically can be traced to the strong ionic nature of the transition metal elements.
Particularly, the open-shell electronic structure of these ions often experiences intra-atomic correlation of 10 eV scale and interatomic correlation of 100 meV scale, both much stronger than room temperature.
Only when constraints of these higher-energy physics are incorporated, the rich \textit{emergent} lower-energy physical behaviors can be properly described.
Specifically, it is essential to identify the dominant valence of the transition metal elements and the leading fluctuation in the charge, spin, or orbital channels.
These Hartree-scale electronic properties would grant researchers a proper starting point to formulate effective models to treat more accurately the delicate and rich lower-energy physics.

However, the geometric frustration of the kagome lattice renders such identification more challenging, both experimentally and theoretically.
This is because such geometric frustration suppresses the classical interatomic spin coupling, allowing the ionic spins to fluctuate quantum mechanically according to their strong coupling to the itinerant carriers with much faster dynamics.
This means that only fast experimental probes with shorter timescale can detect such a rapidly fluctuating local electronic structure.
For example, such rapidly fluctuating local moments in stoichiometricric homogeneous compound would be beyond detection of the muon spin rotation or relaxation ($\mu$SR) technique~\cite{Kenney2021Absence}.
Similarly, the typical experimental temperature would be too low to overcome the itinerant-induced correlation to reveal the entropy-driven Curie-Weiss behavior of local moments in this material.
This also poses a great challenge to typical theoretical treatments of many-body physics~\cite{chao1977kinetic,White2002}, which heavily rely on being able to faithfully ``integrate out'' (not throw out) faster dynamics, an extremely difficult task when the coupling is strong.

Here, we take on this essential task and study the higher-energy electronic structure.
To adhere to the geometrical frustration of the system, we first investigate the realistic local ionic electronic structure of RbV$_3$Sb$_5$ and RbTi$_3$Bi$_5$ in \textit{Curie-paramagnetic} states~\cite{Jiang2022Variation, Jiang2023Pressure} that host disordered \textit{noncollinear} spin orientations of the transition metal ions.
Our results indicate that at the Hartree scale, the strong intra-atomic Coulomb repulsion strongly suppresses charge fluctuation involving the transitional metal ions, resulting in a rather well defined $2+$ valence for both V and Ti.
This implies a rather similar itinerant carrier density residing in the ligands in these two families, except for a different size of the fluctuating ionic spin moments that the carriers couple to.
This result, while still too high in the energy scale to draw a definite conclusion for low-energy observables, supports the observation of similar physical behaviors in these two seemingly distinct families.
Particularly, the experimentally observed CDW~\cite{Kang2022Twofold, Chen2021Roton, Khasanov2022Time-reversal, Jiang2021Unconventional, Xiang2021Twofold, Ratcliff2021Coherent, Uykur2022Optical, Ortiz2021Fermi, Li2022Coexistence, Lou2022Charge-Density-Wave-Induced,Li2021Observation, Luo2022Electronic, Li2022Rotation, Nie2022Charge-density-wave-driven, Wu2022Charge, Yu2021Evidence, Park2021Electronic, Lin2021Complex, Denner2021Analysis, Tan2021Charge, Christensen2021Theory, Setty2021Electron, Feng2021Low-energy, Zhou2022Chern, Tazai2022Mechanism, Subedi2022Hexagonal-to-base-centered-orthorhombic, Liu2021Charge-Density-Wave-Induced, Shumiya2021Intrinsic, Wang2021Charge, Feng2021Chiral, Wu2021Nature, Mielke2022Time-reversal, Liang2021Three-Dimensional, Zheng2022Emergent, Wang2023Structure, Wang2021Distinctive, Yu2021Concurrence, Zhou2021Origin, Ge2023Anharmonic, Zhao2021Cascade,Guguchia2023Tunable, Gupta2022Two, Li2022No}, broken time-reversal symmetry (TRS)~\cite{Khasanov2022Time-reversal, Hu2022Time-reversal, Mielke2022Time-reversal,Jiang2021Unconventional,Shumiya2021Intrinsic,Wang2021Charge,Feng2021Chiral,Denner2021Analysis,Lin2021Complex,Wu2021Nature, Setty2021Electron, Xu2022Three, Asaba2024Evidence, Yu2021Evidence, Yu2021Concurrence, Yang2020Giant, Mi2022Multiband, Guo2022Switchable, Zhou2022Anomalous, Wang2021Electronic, Guguchia2023Tunable}, and nematicity~\cite{Li2022Rotation,Zhao2021Cascade,Miao2021Geometry,Ratcliff2021Coherent,Wenzel2022Optical, Xu2022Three, Asaba2024Evidence, Guo2024Correlated, Wu2022Simultaneous} are naturally among the potential lower-energy states in such spin-fermion coupled systems.

We then experimentally verify the existence of these rapidly fluctuating local moments in Sn-doped $\rm {RbV_3Sb_5}$ compounds.
The introduction of nonmagnetic Sn serves as a local impurity that disrupts the perfect geometric frustration in nearby lattices.
It allows local spin correlation to establish more stable local spin configurations that are detectable by experimental probes due to stronger magnetic hyperfine coupling and a slower fluctuation timescale.
The muon Knight shift and static magnetic susceptibility in RbV$_3$Sb$_{5-x}$Sn$_x$ show a clear Curie-Weiss behavior with magnetic moments density proportional to the doping level.
Moreover, longitudinal field (LF) $\mu$SR measurements revealed spin fluctuations with the temperature-independent muon spin relaxation rate up to 30 K, similar to the behavior observed for frustrated magnets~\cite{Sarkar2019Quantum, Li2016Muon}. This relaxation is associated with a low-energy tail of the fluctuating V moments close to Sn atoms and other impurities.
Our study established a higher-energy picture for the electronic structure common for these two families of kagome superconductors, containing carriers of similar density that couple strongly to (and gain correlation effects from) the rapidly fluctuating ionic spin moments of slightly different sizes between the families.
These results suggest a paradigm change from the current carrier-only picture in the field and offer a foundation for the future development of a proper lower-energy description of the rich unorthodox behaviors observed in these exciting materials of intense current research interest.

\begin{figure*}
\centering
\includegraphics[width=0.75\textwidth]{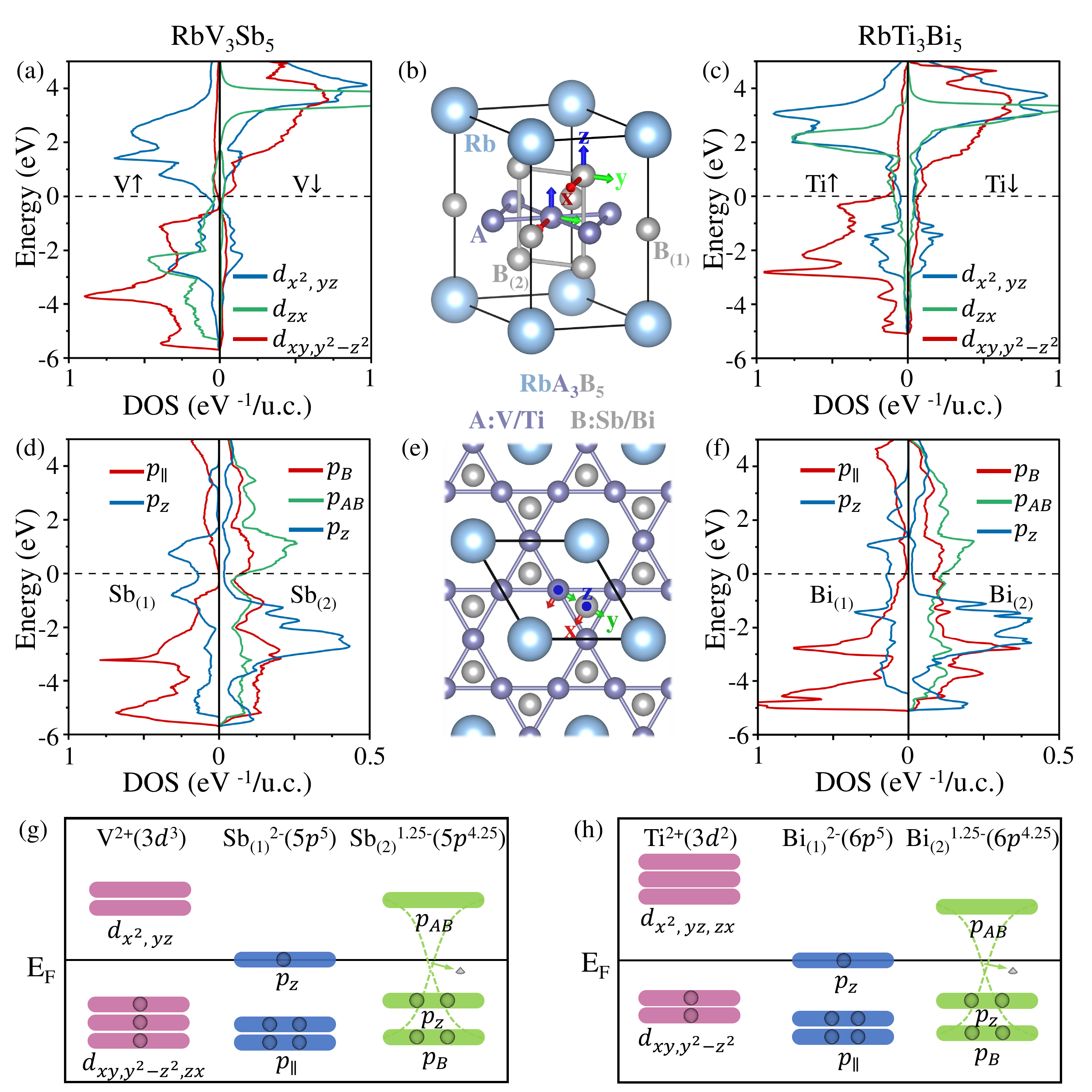}
\caption{ 
Hartree-scale analysis of ionic valence and magnetic moment in kagome metals.
Local orbital-specific density of states (DOS) in noncollinear Curie-paramagnetic phase of RbV$_3$Sb$_5$ for (a) spin majority V$\uparrow$ and minority V$\downarrow$ channels of V indicate a V$^{2+}$ ion with an effective spin-$\frac{3}{2}$ magnetic moment.
The same for (d) in-plane Sb$_{(1)}$ contains a half-filled $p_z$ orbital, corresponding to $\sim1$ itinerant hole carrier.
In contrast, out-of-plane Sb$_{(2)}$ atoms show a strong covalency with $\sim 4$ eV bonding (B) antibonding (AB) splitting (containing $\sim1/4$ itinerant electron carriers).
Note the different horizontal axis scale between left- and right-hand panels for clarify in (d).
These findings are summarized in (g).
Panels (c) and (f) give a similar result for RbTi$_3$Bi$_5$, except a smaller local moment of spin 1 for Ti$^{2+}$, as summarized in (h).
Panels (b) and (e) illustrate the lattice structure and the local coordinate adapted to label local orbitals.
}
\vspace{-0.5cm}
\label{fig1}
\end{figure*}

\section{Theoretical Results}

The key scientific questions of interest in this study are related to Hartree-scale \textit{short-range} physics, such as ionic valence, location of carriers, and their potentially induced interatomic correlations.
Since these higher-energy short-range physics are insensitive to the lower-energy long-range coherence, an informative quantity to investigate is the \textit{atomic orbital-specific} density of states (DOS).
(In contrast, the commonly studied band structure is more sensitive to long-wavelength coherent features, instead of the short-range physics of interest here.)
Still, an accurate investigation is rather challenging due to the frustrated lattice geometry of these materials.
This is because the long-range coherent features, such as the magnetic order, are suppressed by the geometric frustration (as found experimentally), such that short-range physics, for example large ionic moments with strong interatomic coupling, would be easily \textit{masked} at long range and difficult to detect with long-wavelength experimental and theoretical probes.

To this end, we study the ensemble average of atomic orbital-specific DOS along the \textit{internal} spin orientation through simulation of one-body spectral function in a \textit{noncollinear Curie-paramagnetic phase\rm{~\cite{Jiang2022Variation,Jiang2023Pressure}} without artificial magnetic order}, in accordance to the lack of magnetic order in real materials.
This approach allows short-range physics, such as large ionic local moments (if the system prefers), in the absence of long-range coherent magnetic order.
In contrast, typical theoretical studies via density-functional theory (DFT)~\cite{Kang2022Twofold, Hu2022Rich, Ortiz2019New} or perturbation theory~\cite{Watkin2023Fidelity} would not allow such local features without a long-range order. Similarly, studies employing dynamical mean-field theory (DMFT)~\cite{Zhao2021Electronic, Liu2022Weak}, while more accurate on intra-atomic correlation, cannot incorporate itinerant carriers' interatomic multiple scattering processes captured in our approach.
For a comparison between our self-consistent Hartree-Fock approximation and the LDA+DMFT (where LDA is local density approximation) calculation, see Fig.~\ref{figED_DOS} in the Appendix.

The resulting local atomic orbital-specific DOS shown in Figs.~\ref{fig1}(a) and \ref{fig1}(d) allowed us to construct a picture on the dominant charge-valence profile of $\rm {RbV_3Sb_5}$ as summarized in Fig.~\ref{fig1}(g).
Figure~\ref{fig1}(a) shows that despite the lack of long-range ordering, V ions still have a rather well-defined 2+ ($d^3$) valence, with three electrons of the same spin occupying the (red and green) $t_{2g}$ orbitals, forming a spin-$\frac{3}{2}$ ionic moment.
The robustness of effective $d^3$ high-spin structure is further verified under various long-range magnetic configurations [see Supplemental Material (SM)~\cite{supplementary}] and after the introduction of a dilute Sn impurity, as shown in Fig.~\ref{figs5}.
Interestingly, in the left-hand panel of Fig.~\ref{fig1}(d) Sb$_{(1)}$ atoms located in the same layer as V do not have the simple ionic 3- ($p^6$) valence, but instead acquire only 2- ($p^5$) valence.
Correspondingly, the highly mobile (blue) $p_z$ orbitals with large $\sim4$ eV bandwidth are nearly half filled, each hosting one electron carrier on average.

The most surprising is perhaps the Sb$_{(2)}$ atoms located above and below the V layers shown in the right-hand panel of Fig.~\ref{fig1}(d).
The very short distance between Sb$_{(2)}$ atoms within each layer promotes a strong covalency with a bonding- (red) antibonding (green) splitting of an order of $\sim4$ eV.
Consequently, only rather low carrier density (and DOS) associated with them are present near the Fermi level, corresponding to only $\sim\frac{1}{4}$ electron carriers per Sb$_{(2)}$ atom in its $p_\parallel$ orbitals.

Unexpectedly, these high-energy valence profiles are \textit{nearly identical} in $\rm {RbTi_3Bi_5}$, despite the lower electron count (three less electrons per formula unit).
Indeed, following the same analysis above, one finds from Fig.~\ref{fig1}(f) the same valence profile for Bi in $\rm {RbTi_3Bi_5}$ as that for Sb in $\rm {RbV_3Sb_5}$: one electron carrier per Bi$_{(1)}$ in its $p_z$ orbital and $\sim\frac{1}{4}$ electron carrier per Bi$_{(2)}$ in its $p_\parallel$ orbitals.
Instead of shifting the Fermi energy and significantly modifying the itinerant carrier density, the lower electron count of this material is fully absorbed by having one less electron in the $t_{2g}$ orbitals of each Ti ion.
That is, Ti ions acquire the same 2+ valence in $\rm {RbTi_3Bi_5}$ as V in $\rm {RbV_3Sb_5}$, thus hosting a spin-1 ionic moment.
Therefore, from the above Hartree-scale analysis, these two materials have nearly identical valence profiles and similar itinerant carrier density, with only a difference in the size of magnetic moments of the transition metal ions.
(Of course, at lower energy scale, correlation with local moments of different sizes in these materials would naturally lead to distinct itinerant carriers'  long-range dynamics, as manifested in their energy-momentum dispersion and Fermi surfaces.)

Such a robust 2+ valence of V and Ti has an interesting microscopic origin, namely a strong kinetic driven covalency within the ligand layers in these materials, besides the typical large intra-atomic repulsion.
Specifically, Figs.~\ref{fig1}(d) and \ref{fig1}(e) show a large $\sim4$ eV splitting between the red bonding ($p_B$) and green antibonding ($p_{AB}$) $p$ orbitals in one of such in-plane bond.
Such a large energy scale of covalency strongly constrains their potential charge fluctuation and effectively fixes roughly 2 holes in each Sb$_{(2)}$ or Bi$_{(2)}$, which thus in turn helps maintain the robust 2+ valence of V and Ti.
[More precisely, as illustrated in Figs.~\ref{fig1}(g) and \ref{fig1}(h), these orbitals are topologically dictated to cross the Fermi energy with a Dirac-like dispersion~\cite{supplementary}, hosting a low $<\frac{1}{4}$ carrier that makes the electron count $\sim4.25$ per Sb$_{(2)}$ or Bi$_{(2)}$.]

Again, the key difference between $\rm {RbTi_3Bi_5}$ and $\rm {RbV_3Sb_5}$ is the size of magnetic moments. 
Since the latter has a larger ionic spin, presumably the correlation effects generated by its coupling to the itinerant carriers is also stronger.
This is phenomenologically consistent with the stronger tendency toward superconductivity in the latter.

\begin{figure}
\centering
\includegraphics[width=0.95\columnwidth]{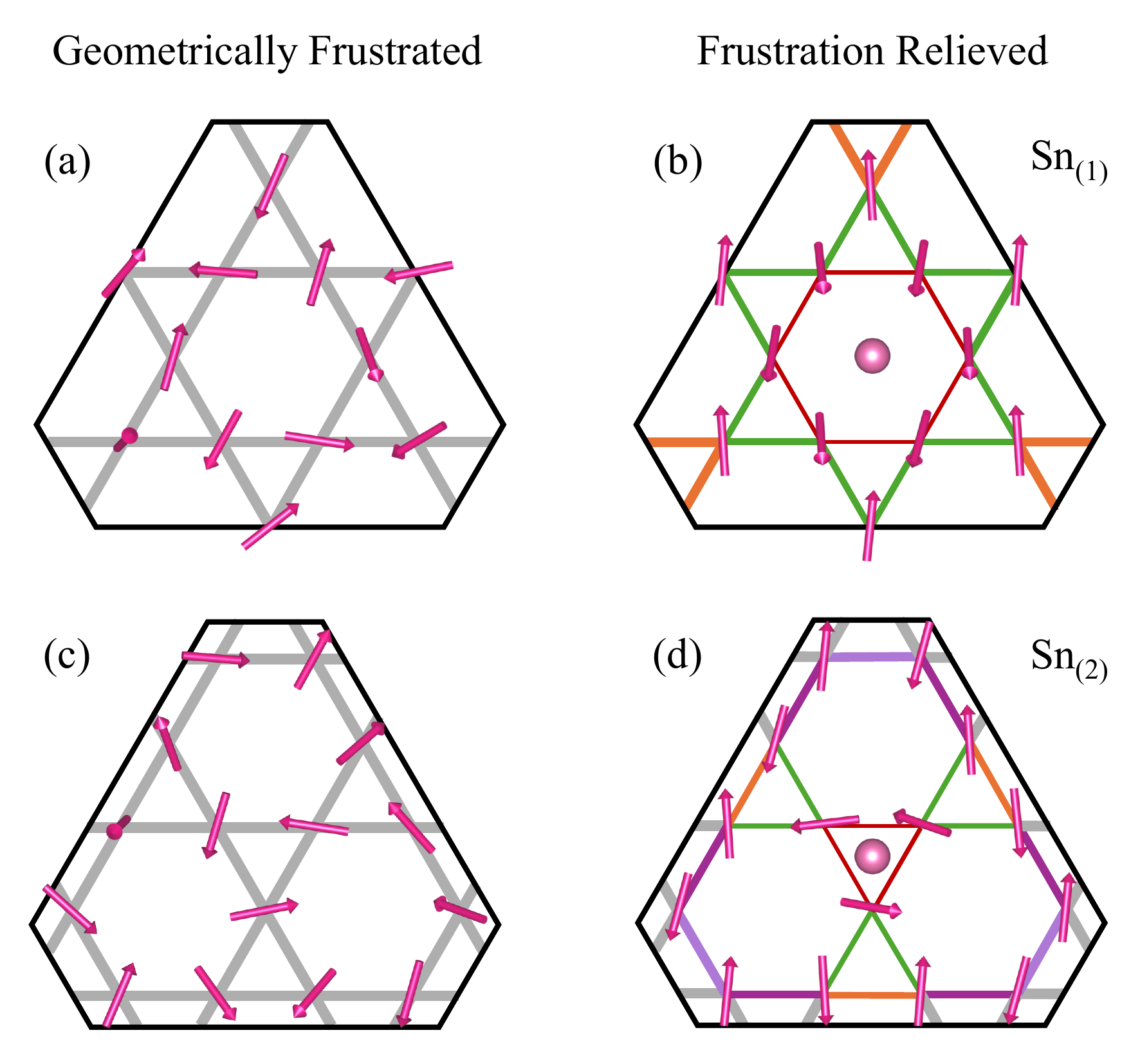}
\caption{
Sketch of relieving geometric frustration to reveal local magnetic moments in kagome metals.
(a), (c) In geometrically frustrated kagome lattice, nearest neighboring antiferromagnetic couplings are completely frustrated, such that classical local moments are energetically free to orient in arbitrary directions.
In kagome metals, coupling to itinerant carriers would fluctuate the moments too fast for typical experiments.
(b), (d) Upon substituting Sb$_{(1)}$ and Sb$_{(2)}$ by (nonmagnetic) Sn impurities, the perfect geometric frustration is relieved, allowing correlation of various strengths (denoted by different colors) between nearest neighboring bonds, and associated with it locally correlated spin structures with slow fluctuation detectable by experiments.
}
\vspace{-0.5cm}
\label{fig2}
\end{figure}

\section{Experimental Confirmation}

Experimental confirmation of the above theoretical findings, particularly the essential large ionic spin, is unfortunately quite challenging.
Because of the lack of a long-range order in geometrically frustrated kagome lattice, long wavelength probes are often insensitive to the local ionic moments.
Furthermore, coupling to the fast kinetic processes of itinerant carriers would fluctuate the ionic spins in a very short timescale, rendering most experimental probes~\cite{Ortiz2019New,Kenney2021Absence} insensitive to their rapid dynamics.

To overcome such difficulty, we attempt to locally slow down the fluctuation and increase the magnetic hyperfine coupling by relieving the geometric frustration through systematic introduction of nonmagnetic Sn impurities to the samples.
The choice of Sn, right next to Sb in the periodic table, is to introduce \textit{minimal} chemical difference (cf. Table~\ref{tabs1}) to relieve the perfect geometric frustration without risking localizing itinerant carriers.
As shown in Fig.~\ref{fig2}, near an introduced impurity, the originally perfect local geometric frustration would be broken, thus allowing nearby ionic spins to correlate better with each other within a small energy scale.
In addition, the presence of impurity also would scatter the itinerant carriers and in turn weaken the quantum coherence in the carrier-induced fluctuation.
Together, both effects would slow the timescale of the fluctuation of the nearby ionic spins and open the opportunity to observe the local magnetic moment via slower probes, as found in other materials~\cite{Maryasin2013Triangular,Ellen2022Diluting,Vojta2000Quantum,Grinenko2011As}.
Naturally, it is necessary to demonstrate a clear doping level dependence to confirm the effectiveness of this proposed mechanism.

It is important to note that a low level of Sn substitution cannot produce local moments similar to that of magnetic V$^{2+}$ ions.
First, in comparison with their interatomic kinetic energy, the intra-atomic interaction of the spatially extended 5$p$ orbitals of \textit{ligands} Sn and Sb is insufficient to support ionic local moments.
Similarly, being next to each other in the periodic table, the difference between their orbital energies is insufficient to localize itinerant carriers under the low density of substitution in our sample.
Particularly, the presence of the localization-resisting Dirac dispersion near the Fermi level should forbid Anderson localization in nearby energies~\cite{Mott1987The}.
Indeed, our realistic calculation (cf. the Appendix) found no ionic Sn moment or localization of itinerant carriers.

\subsection{Magnetic susceptibility}

\begin{figure}[!ht]
\centering
\includegraphics[width=\columnwidth]{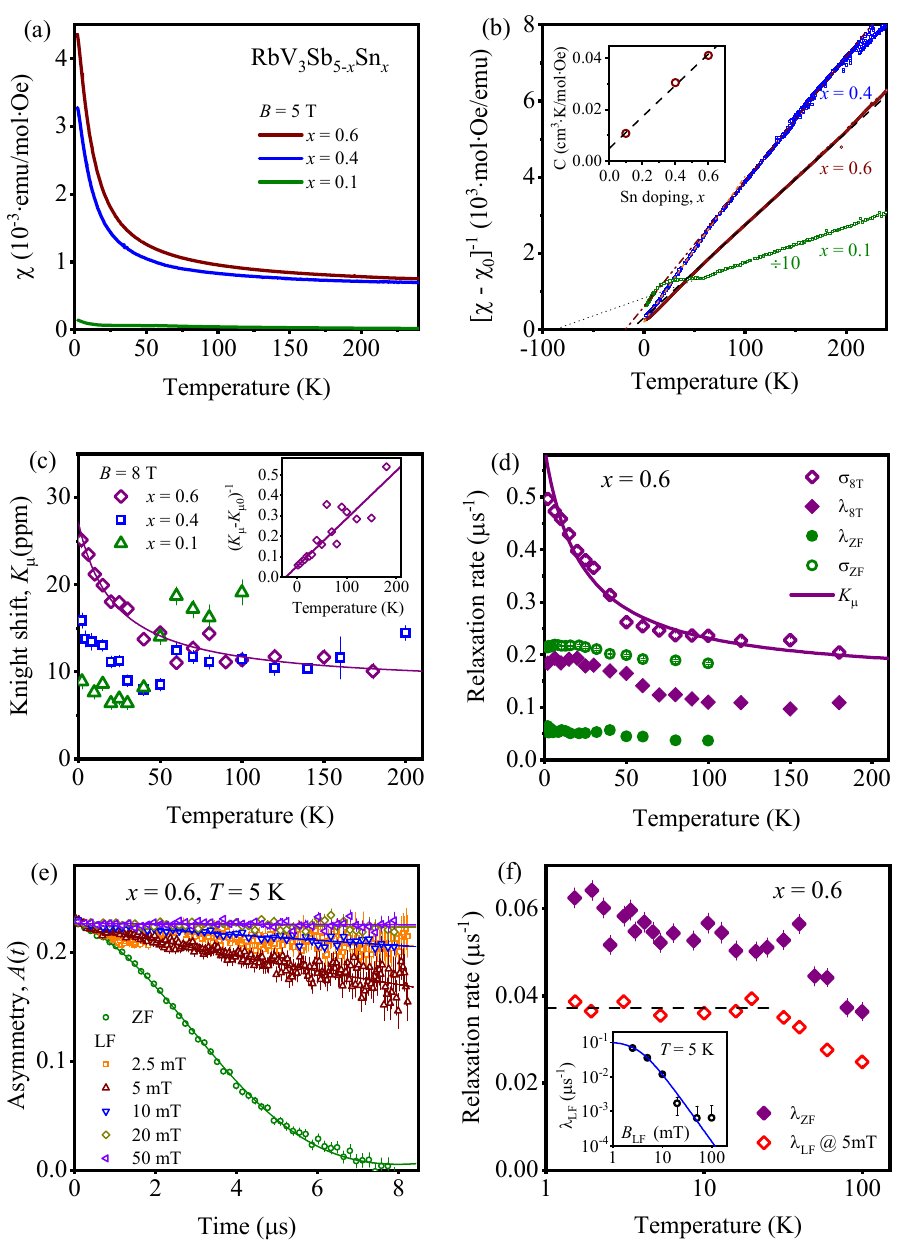}
\caption{Detection of local moments and spin fluctuations in RbV$_3$Sb$_{5-x}$Sn$_x$.
(a) Temperature dependence of the magnetic susceptibility $\chi = M/B$ for polycrystalline samples with different Sn doping levels. The data contain identical extra contributions from the same sample holder. (As a reference, undoped stoichiometric compound data are given in Fig.~\ref{fig_RbV3Sb5_chi}.)
(b) Temperature dependence of the inverse magnetic susceptibility $(\chi-\chi_0)^{-1}$ for data shown in (a); the data for $x = 0.1$ are divided by 10 for clarity. Linear fits are used to define Curie constant $C$. Inset shows $C$ for samples with a different Sn doping level. $C$ increases linearly with the doping level corresponding to an effective moment per Sn atom $p_{\rm eff} \approx 0.7~\mu_{\rm B}$. (c)--(f) $\mu$SR data for the same samples as used for the magnetization measurements.
(c) Temperature dependence of the muon Knight shift $K_{\rm \mu}$ in $B=$ 8 T. The line is a Curie-Weiss fit. Inset shows the inverse Knight shift $(K_{\rm \mu}-K_{{\rm \mu}_0})^{-1}$ for $x = 0.6$.
(d) Temperature dependencies of the muon relaxation rate measured at 8 T and in zero field (ZF). Both Gaussian and Lorentzian contributions to the relaxation rate are shown. The Gaussian relaxation rate at 8 T follows the Knight shift shown in (c).
(e) Asymmetry time spectra measured in ZF and different LF values. The exponential contribution to the relaxation rate survives up to the highest measured fields.
(f) Temperature dependencies of ZF and LF relaxation rate measured at 5 mT. The behavior of the relaxation rate is almost unchanged in the applied field in the whole measured temperature range, indicating a dynamic origin of the relaxation rate. Inset shows the LF dependence of the relaxation rate ($\lambda_{\rm LF}$) obtained from (e). The solid line is a fit with $(a+bB_{\rm LF}^2)^{-1}$ having the form of the standard Redfield equation.}
\vspace{-0.6cm}
\label{fig3}
\end{figure}

The main experimental evidence for the local moments in RbV$_3$Sb$_{5-x}$Sn$_x$ is summarized in Fig.~\ref{fig3}.
Temperature dependence of the static magnetic susceptibility $\chi$ of RbV$_3$Sb$_{5-x}$Sn$_x$ is shown in Fig.~\ref{fig3}(a).
The magnetic susceptibility monotonously increases with Sn substitution level $x$ and for large $x$ shows pronounced Curie-Weiss behavior.
Temperature dependence of $(\chi-\chi_0)^{-1}$ is shown in Fig.~\ref{fig3}(b), where $\chi_0$ is the temperature-independent contribution.
The linear fit with $C^{-1}(T-\Theta)$ results in $C$ values linearly proportional to $x$ [see inset in Fig.~\ref{fig3}(b)], indicating that the amount of local moments is directly proportional to $x$.
This confirms the effectiveness of our attempt to slow down the local spin dynamics near nonmagnetic impurities.

From pure experimental considerations, the above observation is incompatible with magnetism of itinerant carriers near Sn impurities.
First, the doping-induced change of itinerant carrier density is a small fraction of the existing carrier density and thus is unable to account for such strong linear trend.
Second, the default lack of a charge gap in Anderson localization (even if present) implies a Pauli-like susceptibility~\cite{Mott1976ImpurityBand,Freedman1977FermiGlass,Muller1981Pauli,Kamimura1982Theoretical} from \textit{soft} magnetic moments of varying \textit{amplitude}, qualitatively distinct from the entropy-driven Curie-Weiss behavior of \textit{rigid} ionic moments.
(The rigidity of local moments is energetically protected by at least the energy scale of the temperature, 250 K, of our measurement.)

The estimated effective magnetic moment $p_{\rm eff} = \sqrt{3k_{\rm B}C/xN_{\rm A}}\approx 0.7~\mu_{\rm B}$, where $C/x = 0.06$ ${\rm cm^3K/mol_{Sn}}$ is the Curie constant per one Sn atom defined from the linear fit as shown in the inset of Fig.~\ref{fig3}(b), $k_{\rm B}$ is the Boltzmann constant, $N_{\rm A}$ is the Avogadro number, and $\mu_{\rm B}$ is the Bohr magneton.
This value of the local moment $p_{\rm eff}=2\sqrt {S_{\rm eff}[S_{\rm eff}+1]}$ corresponds to a small effective $S_{\rm eff} \approx 1/9$.
This small effective value of the moment compared to the calculated moment for V$^{2+}$ can be explained by the antiferromagnetic orientation of neighboring spins around a Sn impurity, as illustrated in Fig.~\ref{fig2}.
This suggested that even at relatively small concentrations of Sn atoms (in our experiments, the maximum substitution level is about 12$\%$ of Sn atoms), the large sample volume contains magnetic moments.

\subsection{$\mu$SR measurements}

To verify that local moments are the bulk property of the system, we also performed high-resolution muon Knight shift ($K_{\rm \mu}$) measurements in an applied field $B = 8$ T. The analysis of the transversal field (TF) data is explained in the Appendix, and an example of the spectrum is shown in Fig.~\ref{figED_spectra}. In the spectra, we see a single line similar to that observed in undoped high-quality single crystals~\cite{Guguchia2023Tunable}, which excludes secondary magnetic phases' contribution to the measured Knight shift. The obtained temperature dependence of $K_{\rm \mu}$ is shown in Fig.~\ref{fig3}(c) for all doping levels, and $(K_{\rm \mu}-K_{\rm \mu 0})^{-1}$ for the sample with $x = 0.6$ is shown in the inset, where $K_{\rm \mu 0}$ is the temperature-independent contribution. The samples with the lower doping levels show a jumplike feature in $K_{\rm \mu}$ around $T^{*}_2 \sim 50$ K. This temperature was identified in the previous $\mu$SR measurements of stoichiometric RbV$_3$Sb$_5$~\cite{Guguchia2023Tunable}. The jump prevents a reliable Curie-Weiss fitting of the data for $x = 0.1$ and $0.4$. However, it is seen that the upturn in $K_{\rm \mu}$ at low temperatures enhances with the doping increase in accordance with the magnetization measurements. For $x = 0.6$, the jump at $T^{*}_2 \sim 50$ K is not observed, which is consistent with a suppression of the CDW phase~\cite{Oey2022Tuning}.
Nevertheless, there is a change of the muon coupling constant around 50 K for $x$ = 0.6 (see Fig.~\ref{fig_K_vs_chi}).
(The mechanism for this change requires further investigations but is beyond the scope of the present paper, as our extraction of local moments contribution makes use of only susceptibility data above 100 K.)
However, the temperature dependence of the Knight shift is dominated by the paramagnetic contribution due to local moments.
It is seen that $K_{\rm \mu}$ at this doping level follows Curie-Weiss behavior similar to the susceptibility [Fig.~\ref{fig3}(a)], confirming the bulk nature of the paramagnetic contribution.

To study the dynamic properties of the local moments, we analyzed the muon spin relaxation rate obtained from the TF, zero field (ZF), and LF $\mu$SR measurements. Temperature dependence of the TF and ZF relaxation rates are summarized in Fig.~\ref{fig3}(d). The Gaussian component of the TF relaxation rate ($\sigma_{\rm 8T}$) closely follows $K_{\rm \mu}$, indicating that this is a static contribution related to the sample inhomogeneities. The Lorentzian component of the TF relaxation rate ($\lambda_{\rm 8T}$) is small and may have a dynamic contribution~\cite{Grinenko2018Low-temperature}. The ZF relaxation rate is also dominated by Gaussian static contribution ($\sigma_{\rm ZF}$). To verify whether an exponential contribution to the ZF relaxation rate ($\lambda_{\rm ZF}$) is dynamic in origin, we performed LF measurements. The asymmetry plots measured at $T$ = 5 K in different longitudinal magnetic fields are shown in Fig.~\ref{fig3}(e). It is seen that the large Gaussian component of the relaxation rate is completely suppressed by the application of 5 mT. However, despite the small value of $\lambda_{\rm ZF}$, it requires about 50 mT field for complete decoupling of muon spins from internal magnetic fields. The resulting field dependence of the LF muon spin relaxation rate ($\lambda_{\rm LF}$) is shown in the inset of Fig.~\ref{fig3}(f). The field dependence has the standard form expected for fast paramagnetic fluctuations and is observed for many frustrated magnetic systems at this temperature range~\cite{Fujihala2020Gapless, Sarkar2019Quantum, Li2016Muon}. The temperature dependence of $\lambda_{\rm ZF}$ is compared with $\lambda_{\rm LF}$ measured at 5 mT in Fig.~\ref{fig3}(f). It is seen that both relaxation rates have similar temperature dependence with nearly temperature-independent behavior below 30 K. Such behavior is also typical for the frustrated magnetic systems~\cite{Fujihala2020Gapless, Sarkar2019Quantum, Li2016Muon}. The overall behavior of the muon spin relaxation rate in applied longitudinal magnetic fields provides evidence for the presence of paramagnetic fluctuations in the system.

\section{Discussion}

Given the challenge mentioned above in experimental detection of rapidly fluctuating disordered local spins in geometrically frustrated \textit{metallic} systems, the existence of V ionic spins, not surprisingly, has been under heavy debate.
Earlier $\mu$SR measurements performed on high-quality single crystals~\cite{Kenney2021Absence,Mielke2022Time-reversal,Guguchia2023Tunable} did not detect static or fluctuating local moments within the timescale of $10^4$ to $10^{12}$ Hz.
On the other hand, recent $\mu$SR measurements of \textit{undoped} polycrystalline CsV$_3$Sb$_5$ indicated spin fluctuations below 50~K~\cite{Shan2022Muon}, with temperature and field dependences similar to those observed in our data [Fig.~\ref{fig3}(f)].
This new observation likely benefits from a similar (but weaker) relief of geometric frustration as our doped samples.
Indeed, compared to $T_{\rm c}$ of the available single crystals~\cite{Gupta2022Two}, the nearly twofold suppression of $T_{\rm c}$ in the samples of Ref.~\cite{Shan2022Muon} suggests presence of more impurities or dislocations that relieve the geometric frustration and help reveal the V$^{2+}$ local moments.

A lack of data at low temperatures in Ref.~\cite{Shan2022Muon} precludes a detailed comparison of the temperature dependencies of the ZF and LF muon spin relaxation rate. However, there is an indication of a low-temperature plateau, similar to that observed in our experiments [Fig.~\ref{fig3}(f)].
Following the analysis proposed in Ref.~\cite{Aoki2003}, from the field dependence of $\lambda_{\rm LF}\propto[1/(1+\gamma_{\rm \mu}^2B_{\rm LF}^2/\omega_{\rm c}^2)]$ [Fig.~\ref{fig3}(f)], we get $\omega_{\rm c}/\gamma_{\rm \mu}\approx3.5$ mT and a corresponding fluctuation rate $\nu_{\rm c}=\omega_{\rm c}/2\pi\approx0.5$ MHz, where $\gamma_{\rm \mu}$ is the muon gyromagnetic ratio.
The extrapolated relaxation rate to zero field $\lambda_{\rm LF}\approx0.11~\mu s^{-1}$, which corresponds to an average internal field $B_{\rm int}\approx 12$ mT.
These values are comparable to those reported in Ref.~\cite{Shan2022Muon} (note a missing $1/2\pi$ factor in $\nu_{\rm c}$, which makes a difference in the estimated fluctuation rate.)
This comparison indicates a similar nature of the fluctuations in our Sn-doped RbV$_3$Sb$_5$ and polycrystalline CsV$_3$Sb$_5$ samples from Ref.~\cite{Shan2022Muon}, allowing us to conclude that the crystalline defects responsible for suppression of $T_{\rm c}$ in the CsV$_3$Sb$_5$ behave quite similarly to Sn doping in accord with our scenario.

To get further evidence for the universality of the mechanism resulting in the slowing down of spin fluctuations, we investigated the magnetic properties of Sn-doped CsV$_3$Sb$_5$ single crystals. The data are presented in SM~\cite{supplementary}. The doped single crystals pose greater challenges for experimental studies: Submillimeter-sized crystals with substantial Sn doping tend to be inhomogeneous, making quantitative comparisons with homogeneous polycrystalline samples difficult~\cite{Oey2022Tuning}. However, we observed a very clear trend: Sn doping substantially enhances the Curie contribution in CsV$_3$Sb$_5$ single crystals (see Fig.~S5 in SM~\cite{supplementary}).
These allow us to conclude that Sn doping reveals the local moments in both polycrystalline materials and single crystals.

A sample-dependent subdominant Curie-Weiss behavior has been widely observed in magnetic susceptibility~\cite{Ortiz2019New, Ortiz2021superconductivity, Yu2021Concurrence, Wang2021Charge, Fu2021Quantum, Sur2023Optimized}, but often regarded as a consequence of the presence of a minute amount of magnetic impurities, instead of intrinsic local moments in the system.
Also, a small amount of strong magnetic impurities would not be seen in the Knight shift measurements.
The systematic doping dependence to the nonmagnetic Sn impurity in our study invalidates this scenario and instead proves that the magnetic moments are intrinsic to these systems that only become visible to our experimental probe near the impurity.
The same picture also naturally explains the clear observation of doping-dependent Curie-Weiss behavior in Ti-substituted CsV$_3$Sb$_5$ under in-plane magnetic field~\cite{Sur2023Optimized}.
(The strong anisotropy of the electronic structure, as indicated by the superconducting properties~\cite{Ni2021Anisotropic,Gupta2022Microscopic}, might explain the lack of similar systematic doping-dependent Curie-Weiss behavior under out-of-plane magnetic field~\cite{Liu2023Doping}.)
In retrospect, in geometrically frustrated systems, our scenario is more natural and general since crystalline defects (that can serve a similar role as nonmagnetic impurities) are unavoidable in real materials.

Our results are also consistent with recent nuclear magnetic resonance measurements of the spin-lattice relaxation rate, $1/T_1$, of CsV$_3$Sb$_5$~\cite{Song2022Orbital, Luo2022Possible, Zheng2022Emergent} that indicate the presence of spin fluctuations.
Particularly, the first-order nature of the CDW phase transition supports the necessity for contribution from spin fluctuations to the measured $1/T_1$~\cite{Luo2022Possible}.
Furthermore, the pressure independence of the observed Curie-Weiss-like behavior in $1/T_1T$~\cite{Zheng2022Emergent} suggests a rather high-energy scale of the local moments dynamics.

In addition, our theoretical conclusion on the V$^{2+}$ and Ti$^{2+}$ valence is also in agreement with the relative difference of the experimental lattice structures of their corresponding compounds.
For example, let us compare the lengths of the edges, along the $y$ and $z$ directions in Fig.~\ref{fig1}(b), of a plaquette made of the out-of-plane ligand Sb$_{(2)}$ and Bi$_{(2)}$ around V and Ti in RbV$_3$Sb$_5$ and in RbTi$_3$Bi$_5$.
Even though relative to RbV$_3$Sb$_5$ both lengths increase in RbTi$_3$Bi$_5$ due to the larger size of Bi ions; the amount of increase along the $y$ direction $\sim0.174$ \AA~is only half of that along the $z$ direction $\sim0.303$ \AA.
This is in good agreement with a weaker repulsion from electric charge in the $xz$ plane, corresponding to one less electron in the $xz$ orbital of Ti$^{2+}$ ion in our results. (cf. Fig.~\ref{fig1}).

In terms of physical implications, as introduced above, an essential purpose of the above Hartree-scale analysis of the effective ionic charge and valence structure is to establish the many-body correlation and the corresponding reduced phase space that dominate the lower-energy physics.
Governed by such low-energy physics, the static or lower-frequency properties directly connected to experimental observations, such as the superconductivity and CDW formation, can be reliably investigated and properly understood.
For example, upon absorbing the Hartree-scale charge fluctuation involving the transition metal ions and the eV-scale intra-atomic Hund's coupling, the effective low-energy description of these systems would correspond to a generic spin-fermion picture~\cite{Tam2015Itinerancy-Enhanced,Hou2023Chemical,Dagotto2012Anisotropy,Philip2010Orbital,Weng2009Coexistence},
 \begin{equation}
    \begin{split}
        H &= \sum_{ii^\prime m m^\prime \nu}t_{imi^\prime m^\prime}c_{im\nu}^\dagger c_{i^\prime m^\prime \nu} + \sum_{j\neq j^\prime}J_{jj^\prime}\mathbf{S}_j\cdot\mathbf{S}_{j^\prime}\\
          &-\sum_{jim\nu i'm'\nu'}K_{jim\nu i'm'\nu'}\mathbf{S}_j c^\dagger_{im\nu}\bm{\sigma}_{\nu,\nu^\prime}c_{i' m' \nu^\prime},
     \end{split}
    \label{H_eff}
\end{equation}
in which itinerant carriers $c^\dagger_{im\nu}$ of spin $\nu$ in orbital $m$ of ligands located at site $i$ propagate through renormalized \textit{effective} kinetic strength $t$.

Note that in this low-energy effective description, the itinerant carriers are no longer allowed to propagate to the transition metal ions located at site $j$.
Instead, the remaining kinetic processes are renormalized by the suppressed virtual processes through V $t_{2g}$ orbitals.
In addition, the carrier dynamics develops a strong coupling to the ionic spins $\mathbf{S}_j$ through coupling $K_{jim\nu i'm'\nu'}$.
(Here, $\mathbf{\sigma}_{\nu\nu'}$ is the standard vector of Pauli matrices.)
These kinetic and correlation processes are further modulated (and perhaps even compete) with the antiferromagnetic superexchange coupling $J_{jj^\prime}$ between the ionic spins.

Interestingly, recent angle-resolved photoemission spectroscopy~\cite{Lan2026Common} on CsV$_3$Sb$_5$ found clear evidence for sizable band renormalization near the Fermi level, with effective-mass enhancement of order $m^*/m\sim 2$ in some bands.
Similar effective mass was also reported in the quantum oscillation of transport measurements upon suppressing the charge density wave order via pressure~\cite{Wei2024Large}.
Such behavior can be naturally understood in our framework as arising from the \textit{orbital-dependent} renormalization of the kinetic processes due to coupling to the V$^{2+}$ ionic moments.
(In contrast, the very limited number of data points renders the claim of connection to quantum critical fluctuation inconclusive.)

Particularly, due to the geometric frustration of the kagome lattice of $\mathbf{S}_j$ that disables the long-range magnetic order, the larger number of many-ion spin states would \textit{not} energetically spread out by the $J_{jj^\prime}$ scale as in unfrustrated systems.
Consequently, compared to typical doped Mott insulators~\cite{Lee2006Doping}, correlation effects on the itinerant carriers [through the last term in Eq.~\ref{H_eff}] in these systems are therefore expected to be strongly enhanced by such an energy proximity (near degeneracy) of large number of many-ion spin states.
Naturally, such correlation effect is expected to be much more effective to the massive carriers in $\rm Sb_{(1)}$ than the massless carriers in $\rm Sb_{(2)}$.
Combined with their much higher density, one therefore expects the dominant correlation-related physics (and its resulting unusual properties) to be associated with the massive carriers in $\rm Sb_{(1)}$.

This low-energy effective picture offers a natural explanation for the similarly unusual properties of $\rm {RbTi_3Bi_5}$ and $\rm {RbV_3Sb_5}$.
Even though from total electron count of these two compounds differ significantly, by three per chemical unit, at low energy such difference mainly manifests itself through the size of the effective ionic spin, $\frac{3}{2}$ for V and 1 for Ti, and their correlated coupling with the itinerant carriers, rather than the effective carrier densities.
The corresponding similarity in itinerant carrier density nicely explains why the overall itinerant density of states in Fig.~\ref{fig1} and the observed band structures~\cite{Yang2020Giant,Zhou2023Physical,Hu2022Rich} in these two materials do not show a significant (eV-scale) shift against each other.

Specifically for the $\rm {RbV_3Sb_5}$ family, the presence of ionic magnetic moments has significant implications on the current issue of broken time-reversal symmetry indicated by several experiments, including magneto-optic polar Kerr effect studies~\cite{Xu2022Three, Hu2022Time-reversal}, torque measurements~\cite{Asaba2024Evidence}, $\mu$SR measurements~\cite{Mielke2022Time-reversal, Khasanov2022Time-reversal,Yu2021Evidence}, transport measurements~\cite{Guo2022Switchable, Mi2022Multiband, Yu2021Concurrence, Zhou2022Anomalous, Yang2020Giant}, and scanning tunneling microscopy studies~\cite{Jiang2021Unconventional, Guguchia2023Tunable, Wang2021Electronic, Shumiya2021Intrinsic}.
With ionic magnetic moments, such a broken TRS can simply correspond to the presence of a finite ordered moment at long length scale.
Given the large energy of intra-atomic interaction, this scenario relieves the need to generate magnetic moments from itinerant carriers, for example, through the formation of a loop current~\cite{Feng2021Chiral, Hu2022Time-reversal, Park2021Electronic, Tazai2023Charge-loop, Zhou2022Chern, Christensen2022Loop}.
Particularly, the observed in-plane component of magnetic moment in several experiments~\cite{Yu2021Evidence, Guo2022Switchable, Asaba2024Evidence} poses no difficulty for ionic spins, but seems less obvious from in-plane itinerant current.

Our result is also consistent with the observed dramatic enhancement of broken TRS in the charge density wave phase~\cite{Khasanov2022Time-reversal, Hu2022Time-reversal, Mielke2022Time-reversal,Jiang2021Unconventional,Shumiya2021Intrinsic,Wang2021Charge,Feng2021Chiral,Denner2021Analysis,Lin2021Complex,Wu2021Nature, Setty2021Electron, Xu2022Three, Yu2021Evidence, Yu2021Concurrence, Yang2020Giant, Mi2022Multiband, Guo2022Switchable, Zhou2022Anomalous, Wang2021Electronic, Guguchia2023Tunable}.
Thermodynamically stable phase of CDW typically gaps out part of the Fermi surface and in turn reduces the low-energy carrier density.
In spin-fermion systems such as Eq.~\ref{H_eff}, the kinetic motion of itinerant carriers is known~\cite{Anderson1955Considerations, Zener1951Interaction, Yin2010Unified} to promote a better alignment of spins at low carrier density (the so-called ``double exchange'' mechanism~\cite{Zener1951Interaction}).
Alternatively, the observed enhancement of broken TRS can be understood from the suppression of long-range quantum fluctuation~\cite{Tam2015Itinerancy-Enhanced,Hou2023Chemical} of the magnetic order due to reduced carrier density.
Furthermore, in nearly frustrated systems such kinetic mechanism is also known~\cite{Yin2010Unified} to generate nematic structures to balance the kinetic energy and potential energies, in excellent consistency with the observation of enhanced anisotropy in the CDW phase~\cite{Li2022Rotation,Zhao2021Cascade,Miao2021Geometry,Ratcliff2021Coherent,Wenzel2022Optical, Xu2022Three, Guo2024Correlated, Wu2022Simultaneous}.

Our finding of V$^{2+}$ and Ti$^{2+}$ ionic local moments in kagome superconductors should be easily confirmed by experiments of shorter space-time scales, such as the multiplet spectral analysis of the resonant inelastic x-ray scattering or sum-rule analysis of the inelastic neutron scattering, when samples of sufficient quality and size become available.
It can also be theoretically confirmed via rigorous quantum calculations, such as exact diagonalization of local clusters, quantum Monte Carlo simulations, or tensor-network methods.

In short, our discovered ionic moments suggest a paradigm shift from the existing itinerant carrier-only picture to one incorporating strong correlation from local ionic spins.
The fluctuating local moments provide a chemically justified picture capable of unifying seemingly contradictory existing experimental observations.
Furthermore, the associated interatomic and local-itinerant correlations offer a solid ground for further emergence of the observed rich correlated behavior in this new family of superconducting materials.

\section*{Acknowledgments}

We acknowledge Jia-Xin Yin for fruitful discussion and assistance in obtaining single crystals for the research.
This work is supported by the National Natural Science Foundation of China (NSFC) under Grants No. 12274287 and No. 12042507, No. 1231101283, No. 1237040280, and the Innovation Program for Quantum Science and Technology No. 2021ZD0301900.
R. J. acknowledges additional support from a UKRI Future Leaders Fellowship [No. UKRI2083] and from an EPSRC Grant [No. EP/V062654/1].
S. D. W., Y. O., and A. C. S. acknowledge support via the Q-AMASE-i program under Award No. DMR-1906325.
Z. G. acknowledges support from the Swiss National Science Foundation (SNSF) through SNSF Starting Grant No. TMSGI2 211750.
The experiments were carried out at the Swiss Muon Source (S$\mu$S), Paul Scherrer Institute, Villigen, Switzerland.

\section*{Data Availability}

The data supporting this study's findings are available within the article.

\section*{Appendix}

\renewcommand{\theequation}{A\arabic{equation}}
\setcounter{equation}{0}
\setcounter{section}{0}
\setcounter{subsection}{0}
\renewcommand{\thesection}{\arabic{section}}
\renewcommand{\thesubsection}{\alph{subsection}}
\renewcommand{\theHsection}{appendix.\arabic{section}}
\renewcommand{\theHsubsection}{appendix.\arabic{section}.\alph{subsection}}

\section{Theoretical analysis} 

To address the main physical questions of this study, namely, the Hartree-scale charge distribution and intra-atomic correlation in RbV$_3$Sb$_5$ and RbTi$_3$Bi$_5$, we proceed with the following procedure of a controlled calculation: (a) extraction of a generic SU(2) symmetric Hartree-scale Hamiltonian from a series of density-functional theory calculations, (b) evaluation of local orbital-specific density of states in the magnetically unordered \textit{noncollinear} Curie-paramagnetic phase via ensemble averaging the one-body spectral function under the self-consistent Hartree-Fock approximation, and (c) illustration of the robustness of the V$^{2+}$ ionic local moments and the lack of localization of doped hole carriers in RbV$_3$Sb$_{5-x}$Sn$_x$ with the dilute nonmagnetic Sn dopant.

\subsection{Extraction of interacting $H^{\text{(Hartree)}}$ from DFT results}

To extract a generic SU(2) symmetric Hartree-scale Hamiltonian, we followed the previous established procedure~\cite{Jiang2022Variation,Jiang2023Pressure, Jiang2025Pressure} by first performing a large number of DFT calculations under different magnetic structures.
The atomic structures are taken from the experimental refined space group $P6/mmm$ and lattice constants $a=b=5.4715$ \AA~and $c=9.0733$ \AA~for RbV$_3$Sb$_5$~\cite{Ortiz2019New}, and $a=b=5.7731$ \AA~and $c=9.0651$ \AA~for RbTi$_3$Bi$_5$~\cite{Werhahn2022The}, respectively. 
We employ the {\sc Wien2k} implementation~\cite{Blaha1990Full-potential} of the all-electron linearized augmented plane wave method~\cite{Singh2006Planewaves} with LDA+$U$ approximation~\cite{Anisimov1993Density-functional,Liechtenstein1995Density-functional} within the second variational treatment and a ``muffin tin'' radius, $R_\mathrm{MT}=2.5$ a.u., for V and Ti.
The resulting self-consistent Kohn-Sham DFT Hamiltonians are then represented, into a tight-binding form, via the symmetry-respecting Wannier orbitals~\cite{Wei2002Insulating} as effective atomic orbitals.
These Wannier orbitals are constructed to only allow Bloch orbitals within $[-7,20]$ eV around the Fermi energy~\cite{Marzari1997Maximally} and thus have absorbed charge fluctuation into the $s$ orbitals of the ligands Sb or Bi located $\sim 10$ eV below the Fermi energy.

The generic Hamiltonian containing intra-atomic interactions,
\begin{align}
H^{(\mathrm{Hartree})} 
&= \sum_{ii^\prime mm^\prime\nu}t_{ii^\prime mm ^\prime}c^\dagger_{im\nu}c_{i^\prime m^\prime \nu} \label{eq1}\\
&+\frac{1}{2}\sum_{imm^\prime m^{\prime\prime}m^{\prime\prime\prime}\nu\nu^\prime} U_{m m^{\prime\prime} m^\prime m^{\prime\prime\prime}} c^\dagger_{im\nu}c^\dagger_{im^{\prime\prime}\nu^\prime}c_{im^{\prime\prime\prime}\nu^\prime}c_{im^\prime\nu}\nonumber,
\end{align}
is then obtained by demanding it to reproduce each of these Kohn-Sham Hamiltonians under the same corresponding magnetic structure, within the consistent Hartree-Fock treatment of the intra-atomic interaction, under a \textit{single} set of parameters.
Here, $c^\dagger_{im\nu}$ denotes the creation operators of electrons of orbitals $m$ and spin $\nu$ at lattice site $i$, $t_{ii^\prime mm ^\prime}$ the one-body hopping strength, and $U_{m m^{\prime\prime} m^\prime m^{\prime\prime\prime}}$ the intra-atomic Coulomb interaction of transition metal ions.
Specifically, for each attempted value of $U_\mathrm{eff}$ and $J_\mathrm{eff}$ parameters, $U_{m m^{\prime\prime} m^\prime m^{\prime\prime\prime}}$ are assumed to follow the standard Slater integral~\cite{Slater1974Quantum,Liechtenstein1995Density-functional}, while $t_{ii^\prime mm ^\prime}$ are obtained by averaging the DFT Hamiltonian over both spins \textit{after} subtracting out the spin-dependent contributions from the self-consistent Hartree-Fock mean field of the interaction terms~\cite{Anisimov1997First-principles, Wei2006Orbital, Lang2021Strongly}.
The quality of the resulting Hartree-scale Hamiltonian is then examined by comparing its mean-field band structure against that of the LDA+$U$ for all magnetic structures~\cite{Jiang2022Variation,Jiang2023Pressure}.

\begin{figure}
\centering
\vspace{-0.5cm}
\includegraphics[width=\columnwidth]{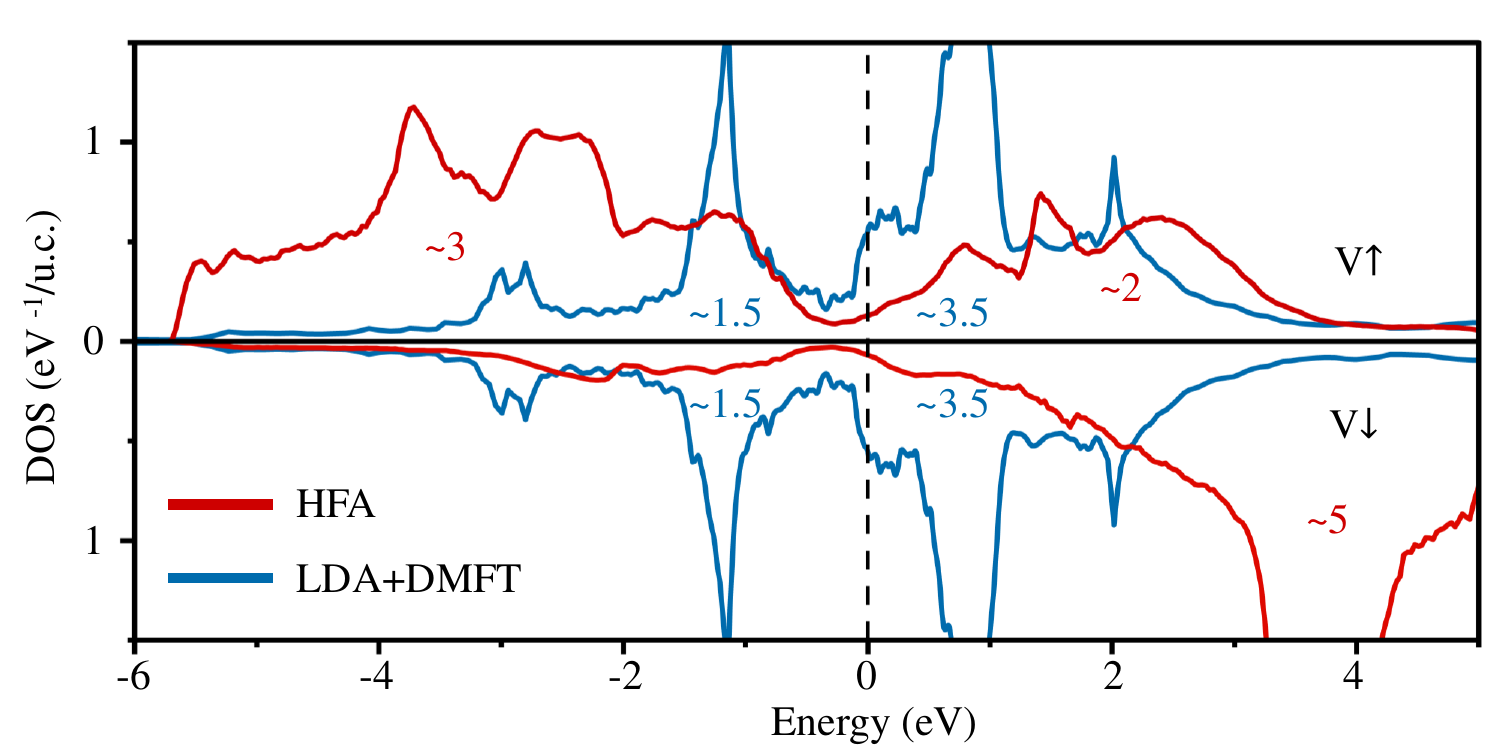}
\caption{Comparing our local orbital-specific density of states (DOS) in the Curie-paramagnetic phase with that obtained from LDA+DMFT method.
Note the direct access to ionic spin structure resulting from our method.
Also note the expected energy splitting associated with the ``bare'' intra-atomic repulsion $\sim5$ eV, in contrast to the reduced energy $\sim2$ eV of the intra-atomic Coulomb repulsion from the LDA+DMFT method.
}
\vspace{-0.3cm}
\label{figED_DOS}
\end{figure}

To more clearly decipher the effective orbital structure and for an easier visualization, in the LDA+$U$ calculations a slightly exaggerated $U-J = 6$ eV is employed inside the muffin tins.
Associated with the abovementioned high-energy hybridization with the ligand $s$ orbitals, this corresponds to a (smaller) realistic $(U_\mathrm{eff}, J_\mathrm{eff})=(4.35, 0.75)$ eV for V and $(4.22, 0.57)$ eV for Ti in the corresponding effective Hamiltonian $H^{(\mathrm{Hartree})}$.
Previously found smooth $U$-dependence of the electronic structure and the effective mass~\cite{Liu2022Weak} in KV$_3$Sb$_5$ for $U\in [2,10]$ eV indicates a single dominant structure insensitive to the value of $U$ and justifies such a convenient choice for qualitative analysis.
Furthermore, we have also confirmed all the qualitative conclusions in the paper under a couple of eV variations of the $U$ parameter~\cite{Jiang2026Anneaing}.

\subsection{Evaluation of local orbital-specific density of states}

To evaluate the local orbital-specific DOS in the magnetically unordered \textit{noncollinear} Curie-paramagnetic phase, we followed the previously established procedure~\cite{Jiang2022Variation,Jiang2023Pressure,Jiang2025Pressure,supplementary} by ensemble averaging the one-body spectral function.
We first set up an ensemble of large supercells, each containing disordered noncollinear spin orientation of transition metal ions with \textit{small} average nearest neighboring spin correlation, $\chi(r) = \sum_i\mathbf{S}_{i+r} \cdot \mathbf{S}_i$.
For each configuration, we evaluate the local orbital-specific DOS along its ionic \textit{spin orientation} through ``unfolded'' one-body spectral function~\cite{Wei2010Unfolding} under the self-consistent Hartree-Fock approximation.
Finally, the resulting orbital-specific DOS is obtained by averaging over the configurations.
Note that a large variation of supercell shapes and orientations is employed to help reduce the artificial periodicity of the supercells~\cite{Jiang2022Variation}.

It is important to note that the above choice of the local orbital-specific DOS is motivated by our scientific question, namely, ionic valence and spin structure in the unordered paramagnetic materials at hand.
As a simple illustration, Fig.~\ref{figED_DOS} compares our resulting local orbital-specific DOS with the local DOS obtained from the state-of-the-art LDA+DMFT methods~\cite{Haule2010Dynamical} under identical parameters.
Note that in the absence of long-range magnetic order, naturally the global DOS from LDA+DMFT is identical between spin orientation.
The more relevant information for this study, such as the effective ionic valence and spin structure, is hidden in the many-body structure embedded in the local impurity solver of DMFT.
In contrast, our local orbital-specific DOS is defined as average over local DOS \textit{along ionic spin orientation} for each ion.
It thus gives distinct DOS for different spin channels, allowing a \textit{direct} visualization of the desired physical information.

One might also notice an interesting distinction in the effect of intra-atomic repulsion between our results and one from LDA+DMFT~\cite{Zhao2021Electronic, Liu2022Weak}.
Specifically, in Fig.~\ref{figED_DOS} the energy splitting between the DOS associated with $d$-orbitals is roughly the same as $U_{m m^{\prime\prime} m^\prime m^{\prime\prime\prime}}$ for our result.
In contrast, the same is heavily ``screened'' in the LDA+DMFT result, by more than a factor of 2.
While such ``screening'' is often observed in DMFT calculations and sometimes agrees with long-wavelength experiments such as angle-resolved photoemission spectroscopy, it should not be effective at the length and energy scale of intra-atomic relevant to the ionic valence and spin structure of interest here.
(Of course, for lower-energy long-wavelength physics, such as quasi-particle spectra of well-defined momentum, a more careful treatment as in DMFT would be necessary to account for the Kondo physics.)

\subsection{Influence of nonmagnetic Sn impurities}

\begin{figure}
\centering
\includegraphics[width=1\columnwidth]{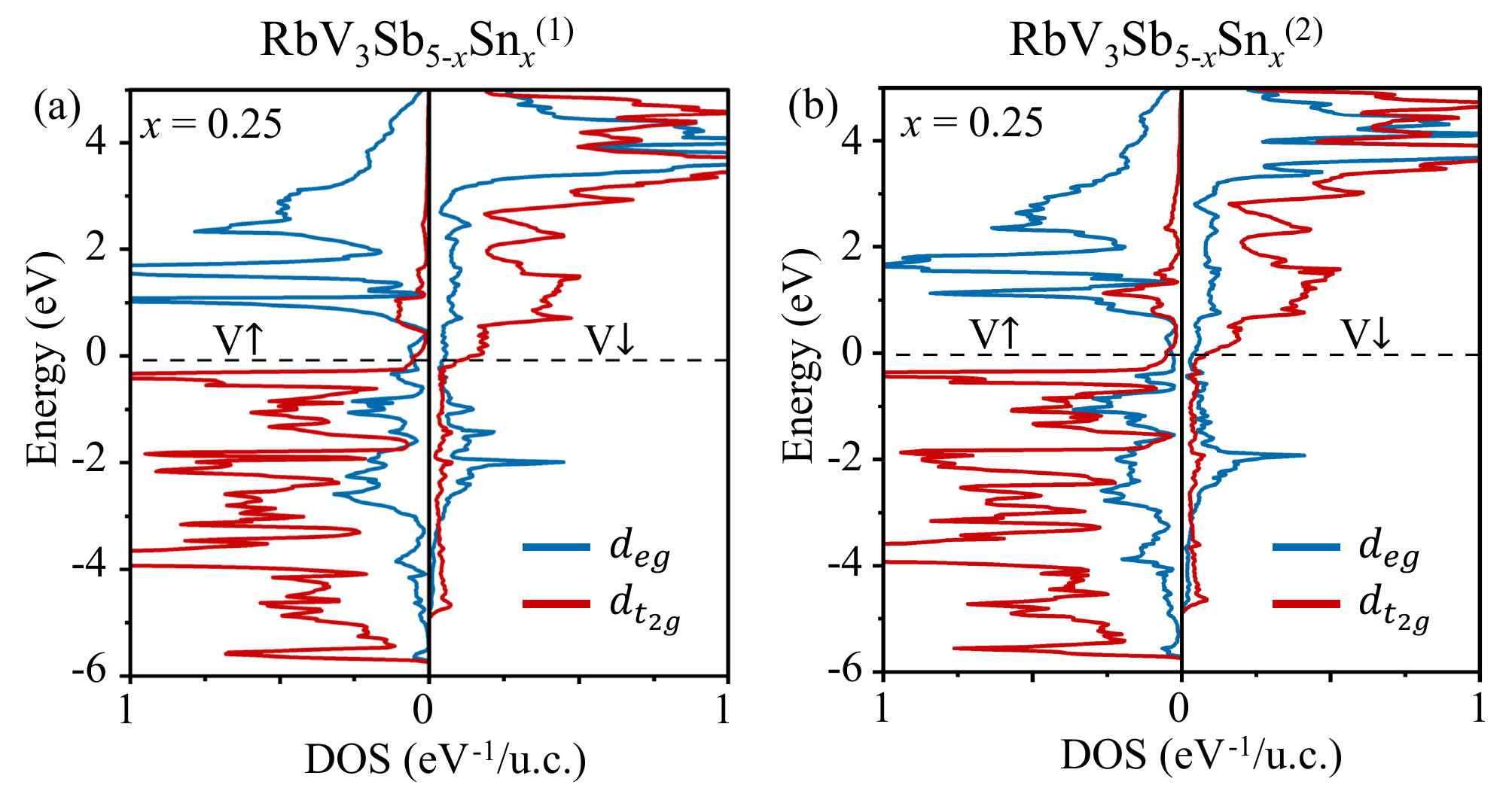}
\vspace{-0.5cm}
\caption{
Robustness of Hartree-scale driven V$^{2+}$ local moments against the dilute nonmagnetic Sn dopant.
Partial density of states of V in the presence of a Sn impurity substituting one of the 20 Sb atoms in a supercell at the (a) Sb$_{(1)}$ site and (b) Sb$_{(2)}$ site (under ferromagnetic order for clarity).
The results display nearly identical lower- and upper-Hubbard structures as the undoped system with an effective $d^3$ high-spin configuration.}
\vspace{-0.3cm}
\label{figs5}
\end{figure}

\begin{figure}
\centering
\includegraphics[width=\columnwidth]{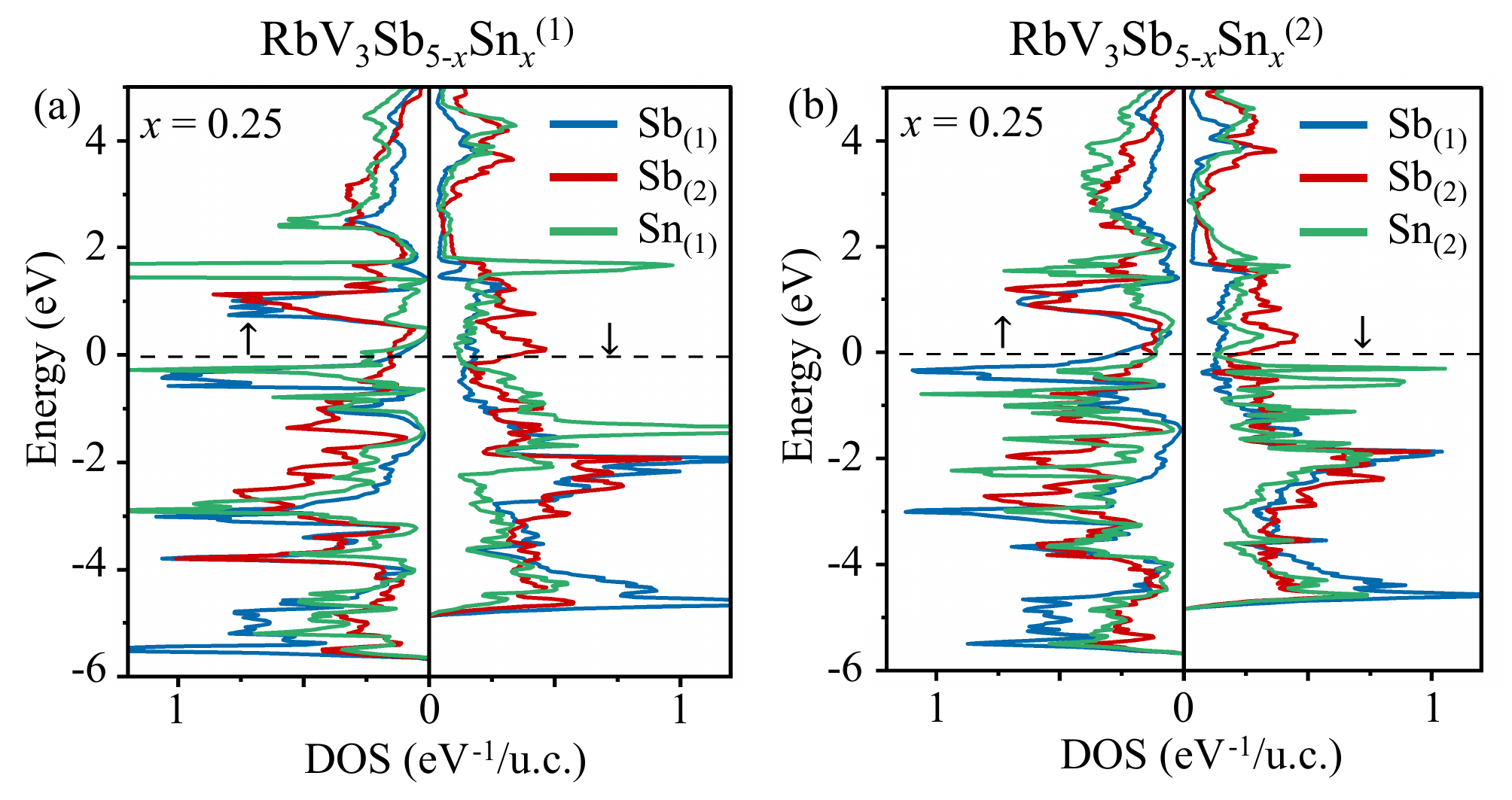}
\vspace{-0.5cm}
\caption{ 
Absence of Sn ionic spin and lack of localized carriers under Sn substitution of Sb.
In great contrast to the V atomic orbitals in Fig.~\ref{figs5}, atomic orbitals of Sn do not display a nearly full spin polarization, reflecting the absence of local ionic moments, even if the system is forced to be ferromagnetic.
Furthermore, as expected from the similar orbital energies of Sn and Sb, at either the (a) Sb$_{(1)}$ or (b) Sb$_{(2)}$ site, their results very closely resemble each other and that of Sb in (c) the pristine compound, showing a large 10-eV energy span and no sign of Anderson localization.
Evidently, the Sn impurities do not induce localized itinerant carriers either.
Therefore, the low level of Sn substitution of Sb does not induce extrinsic local moment.}
\vspace{-0.3cm}
\label{figSbsp}
\end{figure}

Since the ionic electronic structure of V$^{2+}$ is dictated by Hartree-scale intra-atomic interaction, it is naturally insensitive to the sub-eV-scale long-range magnetic ordering and the dilute nonmagnetic weak Sn dopants.
Indeed, the V$^{2+}$ ionic density of states retains its characteristic effective $d^3$ high-spin structure under various long-range magnetic configurations (see SM~\cite{supplementary}) and after the introduction of a single Sn impurity substituting one of the 20 Sb atoms in the supercell, as shown in Fig.~\ref{figs5}.

On the other hand, as ligands, Sn ions are known to be nonmagnetic, as the spatially extended 5$p$ orbitals of Sn dictates that, in comparison with the interatomic kinetic processes, their intra-atomic interaction is insufficient to sustain ionic moments.
This is easily confirmed via a comparison with results of 3$d$ orbitals of magnetic V ions in Fig.~\ref{figs5}, which display a strong spin-polarized 3$d$ orbital density of states, hosting nearly full occupation in the spin majority channel and nearly empty occupation in the minority one.
Indeed, such a strong spin polarization is absent in the Sn 5$p$ density of states in Fig.~\ref{figSbsp}, even if the system is forced to be ferromagnetic.

\begin{table}
\caption{
Influence of the nonmagnetic Sn impurities.
Being next to the Sb, Sn impurities have nearly identical kinetic hopping $t$ parameters to neighboring V-$d_{t_{2g}}$ orbitals, and slightly higher orbital energy $\epsilon$ [still one order of magnitude smaller than their bandwidth (cf. Fig.~\ref{figSbsp})].
Therefore, Sn substitution of Sb cannot induce localization of itinerant carriers, but only alters the nearby superexchange processes and relieves the original perfect geometric frustration.}
\begin{ruledtabular}
\begin{tabular}{ccccc}
 & Sb$_{(1)}$-$p_\parallel$ & Sb$_{(1)}$-$p_z$ & Sn$_{(1)}$-$p_\parallel$& Sn$_{(1)}$-$p_z$  \\ \hline
$\epsilon$ & -1.61 & -0.78 & -0.28 & 0.44 \\ \hline
$t_{pd_{y^2-z^2}}$ & 0.17 & 0 & 0.18 & 0 \\\hline
$t_{pd_{zx}}$ & 0 & 0.66 & 0 & 0.67 \\\hline
$t_{pd_{xy}}$ & 0.53 & 0 & 0.59 & 0 \\\hline
& Sb$_{(2)}$-$p_\parallel$ & Sb$_{(2)}$-$p_z$ & Sn$_{(2)}$-$p_\parallel$& Sn$_{(2)}$-$p_z$  \\ \hline
$\epsilon$ & -0.30 & -1.41 & 1.14 & -0.17 \\ \hline
$t_{pd_{y^2-z^2}}$ & 0.44 & 0.25 & 0.43 & 0.27 \\ \hline
$t_{pd_{zx}}$ & 0.38 & 0 & 0.38 & 0 \\ \hline
$t_{pd_{xy}}$ & 0.31 & 0 & 0.31 & 0 \\
\end{tabular}
\end{ruledtabular}
\vspace{-0.4cm}
\label{tabs1}
\end{table}

Furthermore, Fig.~\ref{figSbsp} shows a very similar density of states for Sn impurities and Sb atoms in the doped and clean samples, all having a large 10-eV energy span and no sign of localized bound state.
Evidently, the very large kinetic energy of Sn and Sb and their similar orbital energy (cf. Table~\ref{tabs1}) does not support Anderson localization of the doped hole carriers.
As a matter of fact, near the Fermi level the band structure contains localization-\textit{resisting} Dirac cones~\cite{supplementary}, coupling to which would disable the possibility of Anderson localized states of similar energy as well.
Therefore, Anderson localization of itinerant carriers is not plausible.

Altogether, these considerations invalidate speculation of extrinsic local moments induced by low level of Sn substitution.
The V$^{2+}$ ionic moments remain the only plausible source for the observed Curie-Weiss behavior.

Therefore, the leading physical effect of the Sn substitution of Sb is to weaken the superexchange process (due to the increase of $\Delta_{pd}\equiv\epsilon_p - \epsilon_d$) and in turn relieve the original perfect geometric frustration.
Estimated via $J\propto \frac{t_{pd}^4}{\Delta^2_{pd}U}$, $J$ is roughly reduced by approximately 25\%--45\%.
Furthermore, since the broken translational symmetry near the Sn dopants would also weaken the itinerant carrier-mediated effective coupling, such reduction on $J$ would locally relieve the geometric frustration and allow nearby ionic spins to correlate better with each other within a small energy scale.
Altogether, these impurity effects would cause the correlated local moments in the vicinity of impurities to fluctuate slowly and behave like free local moments at low temperature, allowing their detection by slow bulk probes such as $\mu$SR measurements.

\section{Experimental analysis}
\label{appendix:experimental}

\begin{figure}[h]
\centering
\includegraphics[width=0.8\columnwidth]{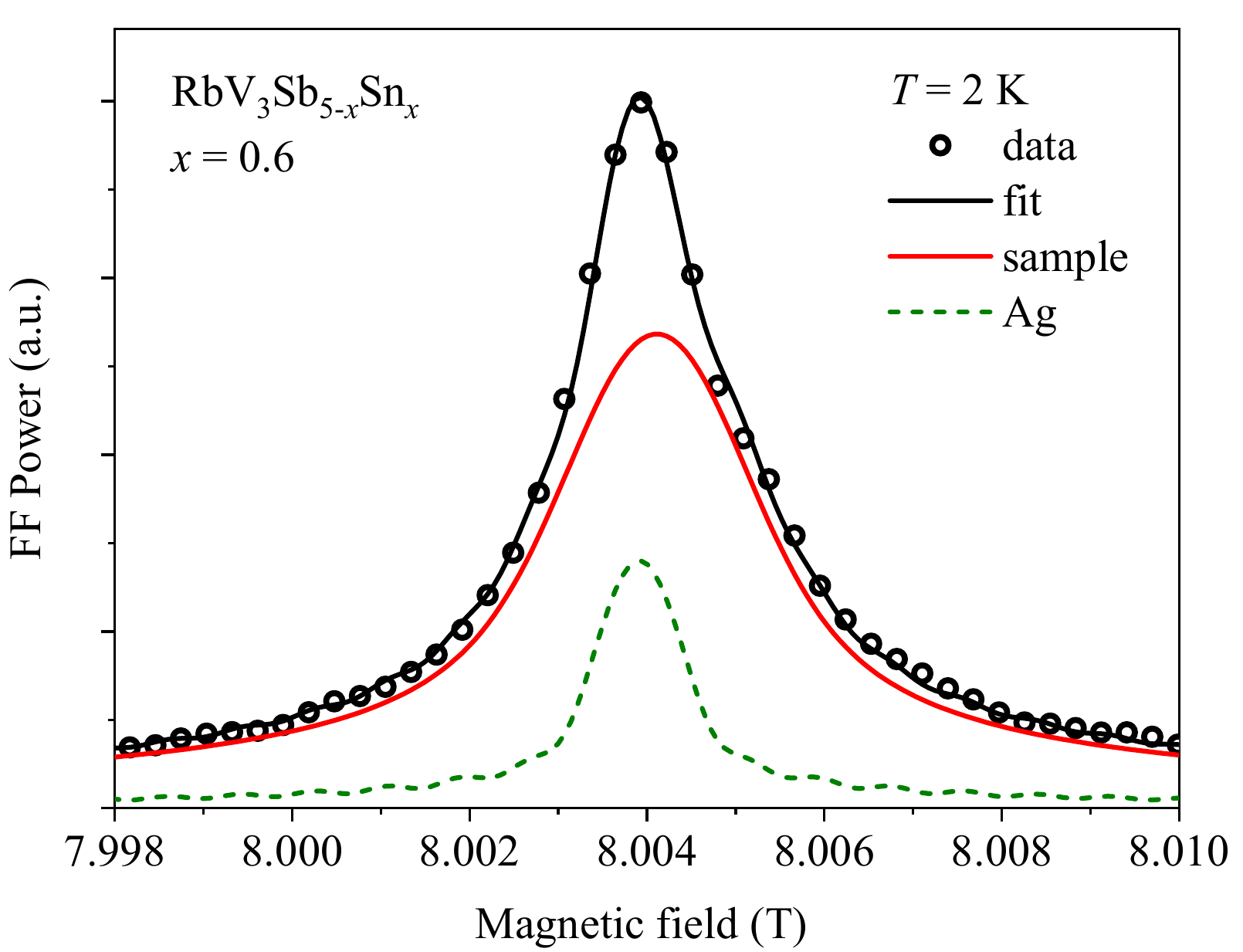}
\caption{High magnetic field spectra of RbV$_3$Sb$_{5-x}$Sn$_x$. An example of the fast Fourier (FF) transform in arbitrary units (a. u.) of the high-transverse-field time spectrum at $T = 2$ K. The fitting curves are described in Sec.~\ref{appendix:experimental} of the Appendix [Eq.~(\ref{eq:tf-fit})].
}
\vspace{-0.3cm}
\label{figED_spectra}
\end{figure}

\begin{figure}
\centering
\includegraphics[width=0.8\columnwidth]{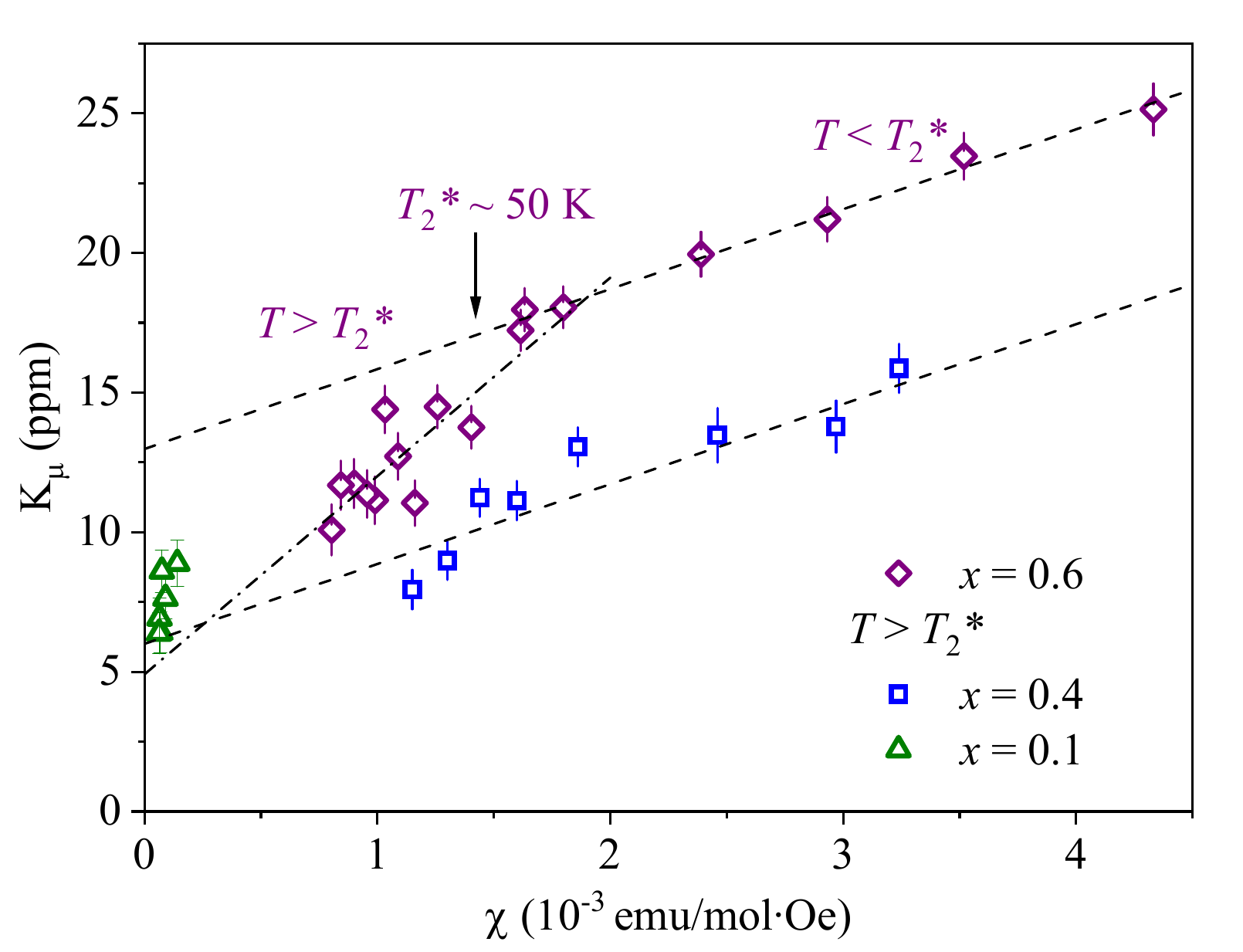}
\caption{Clogston-Jaccarino plot of RbV$_3$Sb$_{5-x}$Sn$_x$. The $\mu$SR Knight shift $K_{\mu}$ versus bulk molar susceptibility $\chi$ for the RbV$_3$Sb$_{5-x}$Sn$_x$ samples with different doping levels. The Knight shift is linear with susceptibility $K_{\mu} = A_{\rm 0}\chi + K_{orb}$, where $A_{\rm 0}$ is the temperature-independent coupling constant and $K_{orb}$ is the orbital contribution to the Knight shift. For $x = 0.1$ and 0.4, the data above $T_2^*$ are not shown for clarity.
For the sample with $x = 0.6$, the jump in the Knight shift at $T_2^*$ is suppressed and only a kink appears.
The observed doping dependence of $A_{\rm 0}$ and $K_{orb}$ is attributed to the suppression of the CDW phase.
}
\label{fig_K_vs_chi}
\end{figure}

\begin{figure}
\centering
\includegraphics[width=0.8\columnwidth]{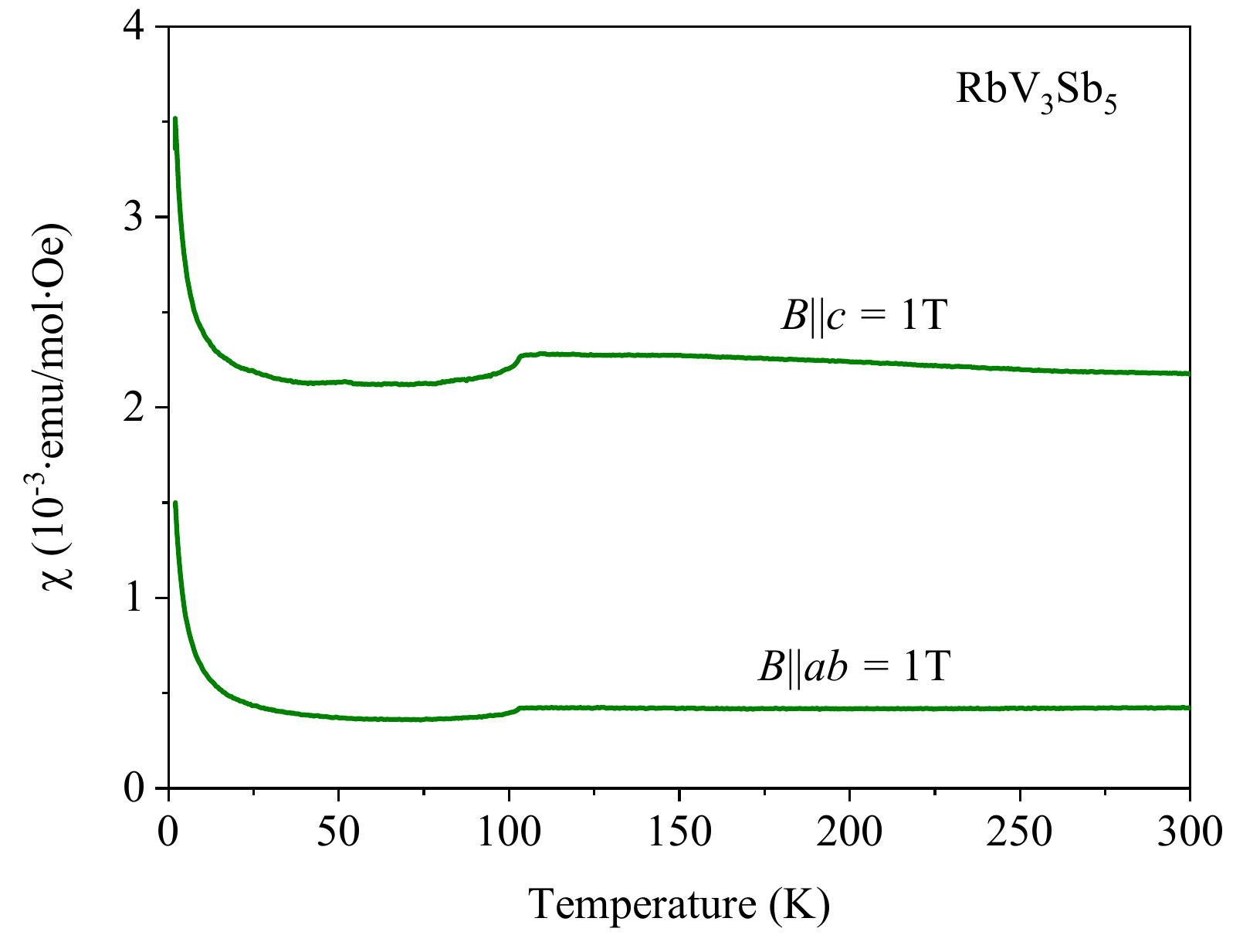}
\caption{Anisotropic magnetic susceptibility of RbV$_{\rm 3}$Sb$_{\rm 5}$. Temperature dependence of the magnetic susceptibility $\chi = M/B$ for RbV$_{\rm 3}$Sb$_{\rm 5}$ single crystal upon removing contribution from the sample holder. Despite a different temperature-independent susceptibility above the $\sim110$ K transition temperature of the charge density wave phase, the overall temperature dependence is qualitatively the same for the two crystallographic directions.}
\label{fig_RbV3Sb5_chi}
\end{figure}

\begin{figure}
\centering
\includegraphics[width=1\columnwidth]{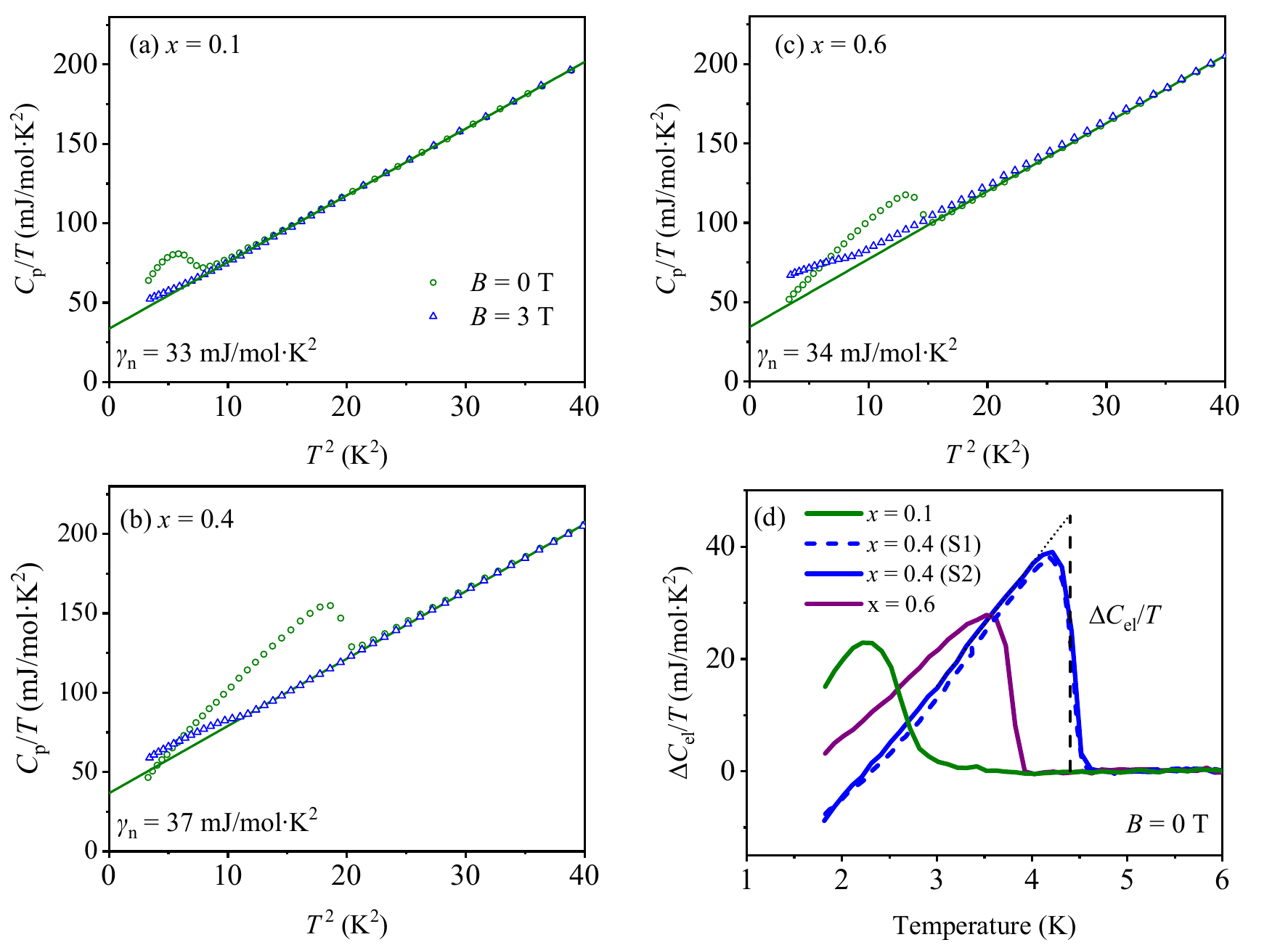}
\caption{Specific heat of RbV$_3$Sb$_{5-x}$Sn$_x$. (a)--(c) Temperature dependencies of the specific heat of the RbV$_3$Sb$_{5-x}$Sn$_x$ samples measured in zero and 3 T magnetic field. (d) Temperature dependence of the electronic-specific heat at low temperatures for samples with different doping levels. The measurements were performed on small pieces cut from the $\mu$SR samples. To demonstrate the homogeneity of the samples, two pieces, S1 and S2, for $x = 0.4$, were measured. The samples show sharp superconducting transitions indicating high crystalline quality and homogeneous doping distribution within the samples. A broader anomaly for $x = 0.1$ is due to the strong doping dependence of $T_{\rm c}$ at this part of the phase diagram~\cite{Oey2022Tuning}.
}
\label{fig_SH}
\end{figure}

\begin{figure}
\centering
\includegraphics[width=0.9\columnwidth]{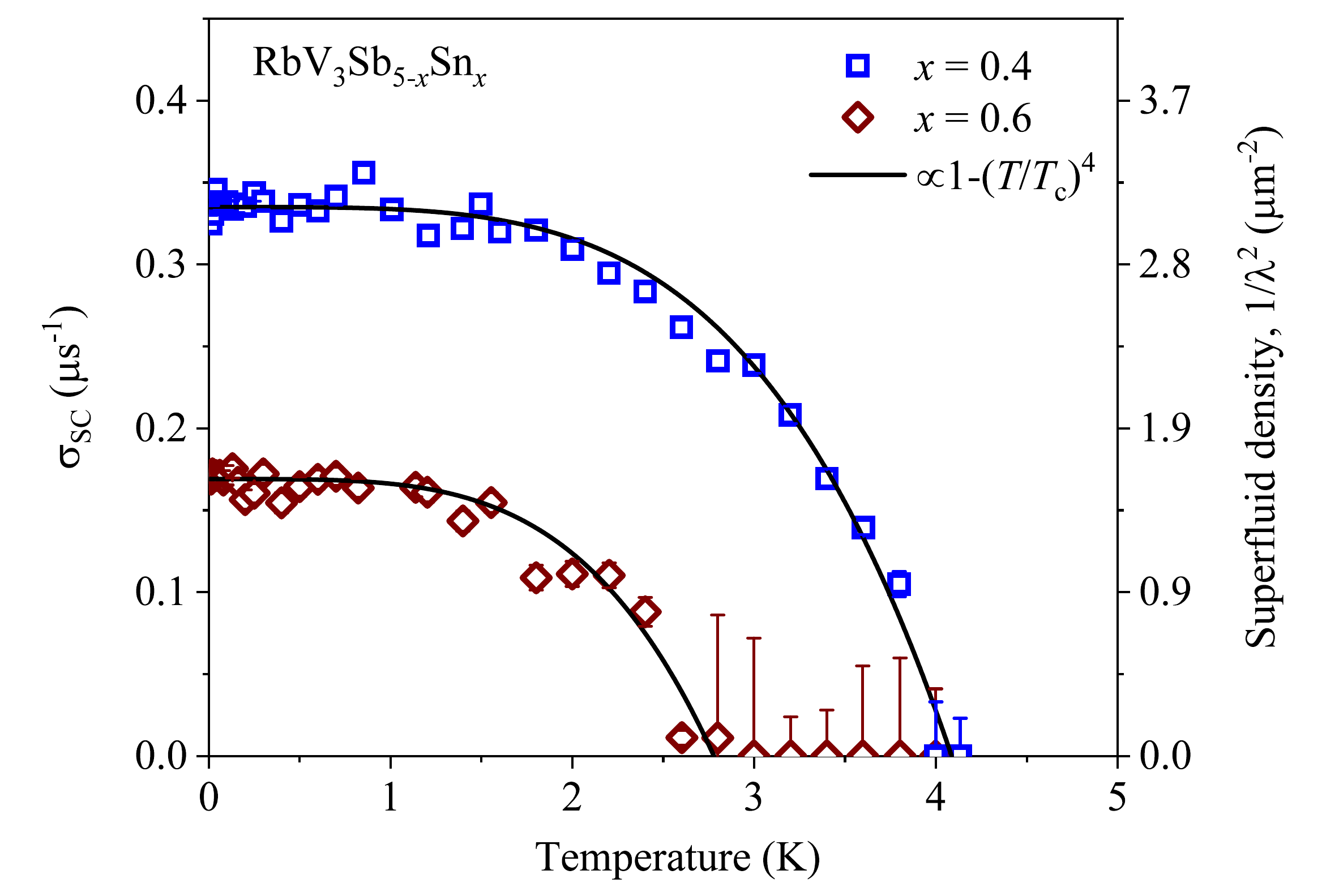}
\caption{Superfluid density of RbV$_3$Sb$_{5-x}$Sn$_x$. Temperature dependencies of the superfluid density of the RbV$_3$Sb$_{5-x}$Sn$_x$ samples measured in 10 mT transverse magnetic field. The superfluid density was estimated using the relation between the magnetic penetration depth $\lambda$ and the second moment $\langle \Delta B^2\rangle\approx (\sigma_{\rm SC}/\gamma_{\mu})^2$ of the internal field distribution, assuming a regular vortex lattice~\cite{Maisuradze2009}. The obtained $\lambda$ should be considered as an approximate quantity since we deal with a polycrystalline sample.}
\label{sigam_sc}
\end{figure}

Single-phase polycrystalline samples of RbV$_3$Sb$_{5-x}$Sn$_x$ were synthesized as described in Ref.~\cite{Oey2022Tuning}. Structural analysis was performed via a Panalytical Empyrean laboratory x-ray powder diffractometer. A Hitachi TM4000Plus scanning electron microscope was used to perform energy-dispersive x-ray spectroscopy. The sharp anomaly in the specific heat at $T_{\rm c}$ (Fig.~\ref{fig_SH}) and the large superfluid density (Fig.~\ref{sigam_sc}), comparable to those measured for nondoped RbV$_3$Sb$_{5}$ single crystals under pressure~\cite{Guguchia2023Tunable}, confirms the high crystalline quality of the samples and homogeneous distribution of the Sn within the sample volume. The slightly smaller value of the jump $\Delta C_{\rm el}/T_{\rm c}\gamma_{\rm n} \approx 1$--$1.2$ compared to the BCS predictions, 1.43, can be attributed to a multiband nature of the superconductivity in this compound, where $\gamma_{\rm n}$ is the normal state electronic specific heat (see Fig.~\ref{fig_SH}).

The magnetic susceptibility data for the reference non-doped RbV$_3$Sb$_5$ single crystal are shown in Fig.~\ref{fig_RbV3Sb5_chi}. The susceptibility is anisotropic, which corresponds to the average susceptibility of a randomly oriented polycrystalline sample given by $\chi_{\rm aver}=\frac{1}{3}\chi_c+\frac{2}{3}\chi_{ab}$. The data also show a noticeable upturn at low temperatures, which in our picture is related to local moments that can stem from several contributions. (i) Disorder effect: some types of disorder locally slows down fast magnetic fluctuations similarly to the Sn effect discussed in the main text. (ii) CDW domain walls: the distortions of the crystal lattice at the domain walls can also release frustrations and induce weak magnetism. Further studies are needed to investigate the possible origin of magnetism in the stoichiometric compound.

The $\mu$SR measurements were performed on pellets with a diameter 8 mm and height 2--3 mm prepared by pressing RbV$_3$Sb$_{5-x}$Sn$_x$ powders.
Zero-field (ZF) and longitudinal-field (LF) measurements were performed using the GPS instrument. The $\mu$SR data were analyzed using the musrfit software package~\cite{Suter2012Musrfit}. The ZF data were analyzed using
 \[
A_\text{fit}(T, t) = A_\text{s} e^{-\lambda_{\rm ZF}t}[\frac{1}{3}+\frac{2}{3}[1-(\sigma_{\rm ZF}t)^2]e^{-\frac{1}{2}(\sigma_{\rm ZF} t)^2}] + A_\text{bkg},
\]
where $\lambda_{\rm ZF}$ and $\sigma_{\rm ZF}$ are muon spin depolarization rates, $A_\text{s}$ sample asymmetry, and $A_\text{bkg}$ is the asymmetry of the background due to muon stopped in material other than the sample. $A_\text{bkg}$ was defined using low-field TF measurements in the superconducting state as proposed in Refs.~\cite{Grinenko2021Split,Grinenko2020Superconductivity}. Because of the large sample size and veto detector system, we found $A_\text{bkg} \sim 0$. For analysis of LF data, we used the same model as for ZF with LF Gaussian Kubo-Toyabe term~\cite{Hayano1979Zero}.

The high-resolution Knight shift measurements at transverse magnetic field $B = 8$ T were performed as proposed in Ref.~\cite{Grinenko2018Low-temperature}. The same samples as in ZF and LF experiments were used for TF measurements using HAL-9500 instrument. The Knight shift is defined relative to an Ag reference, as shown in Fig.~\ref{figED_spectra}. 
To get the background of a reasonable size we intentionally leave the part of the Ag holder uncovered. 
The obtained fraction of the Ag background fits well to one estimated from the area covered by the sample.
Correction for diamagnetic effects was not performed. 
The data were fitted in the time domain using one cosine component for the sample ($A_{\rm s}$) and one for Ag reference ($A_{\rm Ag}$) with two different frequencies. 
In accordance with previous measurements of RbV$_3$Sb$_5$, we described the muon spin depolarization rate using Gaussian ($\sigma_{\rm 8T}$) and Lorentzian ($\lambda_{\rm 8T}$) contributions.
\begin{equation}
\begin{split}
    A(t) &= A_{\rm s}\cos\left(2\pi\nu_{\rm s} t+\frac{\pi\varphi}{180}\right)e^{-\lambda_{\rm 8T}t}e^{-\frac{1}{2}(\sigma_{\rm 8T}t)^2}\\
    &+ A_{\rm Ag}\cos\left(2\pi\nu_{\rm Ag} t+\frac{\pi\varphi}{180}\right)e^{-\lambda_{\rm Ag}t},
\end{split}
\label{eq:tf-fit}
\end{equation} 
In the analysis of the data, Ag contribution was defined at 2 K and fixed for other temperatures. The Ag used for the reference does not show any resolvable temperature dependence in the Knight shift or damping rate~\cite{Grinenko2018Low-temperature,Grinenko2021Split}. The results are summarized in Figs.~\ref{fig3}(c) and \ref{fig3}(d).

To further demonstrate the bulk nature of the local moment contribution to the magnetic susceptibility shown in Fig.~\ref{fig3}, we plotted in Fig.~\ref{fig_K_vs_chi} the muon Knight shift ($K_\mu$) vs magnetic susceptibility (Clogston-Jaccarino plot). The linear correspondence between the Knight shift and the susceptibility for the samples with $x = 0.4$ and 0.6 demonstrates the same origin of the magnetic signal, with a proportionality constant given by the muon hyperfine coupling.

The magnetization measurements were performed in a Quantum Design physical property measurement system using a vibrating-sample magnetometer on pieces of the samples cut from the pellets used for the $\mu$SR experiments. The sample masses were about 30 mg.

\bibliography{main.bib}
\end{document}